\documentclass[aps,pre,preprint,superscriptaddress,amsmath,amssymb]{revtex4-2}
\usepackage{graphicx}
\usepackage{dcolumn}
\usepackage{bm}
\usepackage[mathlines]{lineno}
\usepackage{epsfig}
\usepackage{xcolor,colortbl}
\usepackage{tabularx}
\usepackage{epstopdf}
\usepackage{amsmath,amssymb,amsthm}
\usepackage{bm}
\usepackage{float}
\usepackage{xcolor}
\usepackage{mathtools}
\usepackage{subfig}
\usepackage{hhline}
\usepackage{algcompatible}
\usepackage{newfloat}
\usepackage[justification=RaggedRight]{caption}
\DeclareFloatingEnvironment[
fileext=loa,
listname=List of Algorithms,
name=ALGORITHM,
placement=tbhp,
]{algorithm}

\begin{document}
	\title{T3ST code: Turbulent Transport in Tokamaks via Stochastic Trajectories}
	
	\author{D. I. Palade}
	\email{dragos.palade@inflpr.ro}
	\affiliation{National Institute of Laser, Plasma and Radiation Physics,	M\u{a}gurele, Bucharest, Romania}
	\affiliation{Faculty of Physics, University of Bucharest, Măgurele, Romania}
\author{L. M. Pomarjanschi}
\affiliation{National Institute of Laser, Plasma and Radiation Physics,	M\u{a}gurele, Bucharest, Romania}
\affiliation{Faculty of Physics, University of Bucharest, Măgurele, Romania}

\date{\today}

\begin{abstract}
We introduce the Turbulent Transport in Tokamaks via Stochastic Trajectories (T3ST) code, designed to address the problem of turbulent transport using a statistical approach complementary to gyrokinetics. The code employs test-particle methods to track the dynamics of charged particles in axisymmetric magnetic equilibria, accounting for both turbulence and Coulomb collisions. The turbulence is decoupled from plasma dynamics and represented through a statistical ensemble of synthetic random fields with specified spectral properties. This approach enables T3ST to compute transport coefficients as Lagrangian correlations—orders of magnitude faster than gyrokinetic codes.
\end{abstract}


\maketitle

\section{Introduction}
\label{Section_1}

Nuclear fusion is widely regarded as one of humanity's most promising solutions to address the growing energy demands of the modern world. Despite remarkable advancements over the past 70 years, the realization of viable controlled thermonuclear fusion remains decades away. Among the most advanced experimental configurations are tokamak devices \cite{wesson2011tokamaks,Mazzi2022, Joffrin_2024,Mailloux_2022}, which use strong, axisymmetric toroidal magnetic fields (typically $B\sim 1T$) to confine hot hydrogen plasmas (typically $T\sim 1keV$). However, the geometric complexity and the rich, nonlinear, dynamic phenomena inherent to tokamak plasmas result in significant radial transport, which poses a serious challenge to effective confinement.

Within the hot core of the plasma, transport is primarily neoclassical, driven by collisional processes \cite{Hirshman_1981}. Moving away from the core, small fluctuations extract free energy from plasma gradients, become unstable and evolve into turbulent states, characterized by chaotic fluctuations on microscopic scales \cite{RevModPhys.71.735}. Over the past few decades, considerable progress has been achieved in understanding turbulence-driven anomalous transport in fusion plasmas, largely through state-of-the-art numerical simulations grounded in the gyrokinetic (GK) framework \cite{Brizard_hahm_gyrokinetic}. Nevertheless, a complete understanding of these phenomena remains elusive, in part due to the substantial computational resources required for gyrokinetic simulations, even by current standards.

The neoclassical transport is described by a well-developed theoretical framework, implemented through numerous numerical codes with varying methodologies and capabilities. Among the prominent examples, we highlight NEO \cite{Belli_2008}, NCLASS \cite{10.1063/1.872465}, and ASCOT \cite{HIRVIJOKI20141310}. The turbulent transport is primarily addressed using the gyrokinetic theory \cite{Brizard_hahm_gyrokinetic}. This field also encompasses a vast array of numerical tools, reflecting diverse approaches and modeling capabilities. Notable examples of gyrokinetic codes include the global ORB5 \cite{LANTI2020107072}, gradient-driven codes like GENE \cite{JENKO2000196} and GKW \cite{PEETERS20092650} and flux-driven codes such as GYRO \cite{CANDY2003545}, GYSELA \cite{GRANDGIRARD2006395}, and GT5D \cite{Idomura_2005}. It is worth highlighting the success of quasilinear approaches, particularly the QualiKiz framework \cite{Bourdelle_2016,Stephens_Garbet_2021,10.1063/5.0174643,10.1063/1.2800869}, which trade accuracy for computational speed by linearizing the gyrokinetic equation around a Maxwellian quasi-equilibrium. Finally, we must mention other numerical codes (LOCUST\cite{Ward_2021}, ORBIT\cite{osti_5168290}, PTC\cite{Wang_2021}, etc.) that employ particle tracing and investigate more limited problems: guiding-center trajectories, the dynamics of fast-ions, alpha particles, electrons, etc. 

In this paper, we present the T3ST code (Turbulent Transport in Tokamaks via Stochastic Trajectories, pronounced  "test"). This is a Lagrangian code that evaluates the gyrocenter stochastic trajectories of a test-particle ensemble moving under the influence of tokamak electromagnetic equilibrium, Coulomb collisions and turbulent fluctuations. The latter are not computed self-consistently, as GK does, but generated synthetically through a statistical ensemble of random fields with prescribed spectra. Transport coefficients (or, equivalently, fluxes) are evaluated as Lagrangian averages over the super-ensemble of particles and fields. 

Given the complex and extensive landscape of existing numerical codes, one might question the need to develop a new code, such as T3ST. The motivation lies in addressing a critical gap within the fusion community: the lack of a numerical tool for turbulent transport modeling that combines speed, nonlinear dynamics, and versatility. Compared to existing particle or neoclassical codes, T3ST has the distinct advantage of tackling turbulent transport. In addition to quasilinear methods (such as QualiKiz), T3ST is fully non-linear. However, relative to gyrokinetic and quasilinear approaches, T3ST has a notable limitation: it does not compute explicit turbulent fields but instead generates them synthetically. Paradoxically, this limitation is also one of T3ST's key strengths. By avoiding the need to solve field-matter equations, T3ST achieves computational speeds that are orders of magnitude faster than GK simulations. For instance, low-resolution computations of transport coefficients can be completed in minutes or even seconds on a standard personal CPU with modest specifications (e.g., 2.6 GHz and 6 physical cores). Extensive parametric studies could be conducted on medium-sized computational clusters ($\sim 100$ cores) within a matter of hours. Additionally, T3ST’s synthetic approach offers significant versatility for exploring various physical regimes of interest. Through straightforward manipulation of the turbulent fields, users can easily investigate how specific turbulence characteristics, such as correlation lengths, influence transport - a task that is considerably more challenging to implement within a GK code.

In terms of novelty, T3ST is the refined version of previous, more simplified, codes developed by the authors in the recent years \cite{Palade2021,Palade_2021_W,Palade_2023,Vlad_2021,palade_pom_2022,Palade_2024_scaling} although, the theoretical framework of T3ST was present in literature in various forms for some time. First of all, T3ST is buildt on a well-known statistical approach \cite{balescu2005aspects}: since turbulence is inherently chaotic, one should describe turbulent fields at a statistical level, using a statistical ensemble of random fields that replicate the key properties of real turbulence. Second, T3ST uses test-particles to evaluate transport. This approach has been employed many times before in relation with tokamak plasmas \cite{10.1063/1.1647136,10.1063/1.4844035,PhysRevLett.91.035001,PhysRevLett.88.195004,PhysRevLett.101.095001,10.1063/1.3379471,10.1063/1.1647136} but almost always is buildt on top of gyrokinetic/fluid codes, thus, on a single realistic turbulent state. The use of test-particle methods in conjunction with statistical ensembles has been arround for some time \cite{Vlad_2004}, etc. One notable example of numerical simulations that are quite close in nature to the present work can be found in \cite{10.1063/1.3013453}.

The paper is organized as follows. In Section \ref{Section_2}, we describe the dynamical equations of particle gyrocenters, with emphasis on the representation of geometrical aspects, magnetic fields, turbulence, and collisions. Section \ref{Section_3} discusses numerical details of the implementation of T3ST. Section \ref{Section_4} is reserved for testing the code agains analytical and numerical results. Finally, in the Section \ref{Section_5}, we review conclusions and perspectives.

\section{Theoretical background}
\label{Section_2}

The T3ST code is designed to evaluate local transport coefficients (equivalently, fluxes) in tokamak plasmas using the method of test-particles, also referred to as direct numerical simulation or Monte Carlo \cite{hauff0,hauff1,Palade2021,Palade_2021_W}. The particles are captured at the level of their gyrocenters, which are driven by three primary components of motion: neoclassical equilibrium forces, which account for typical magnetic drifts, centrifugal effects, toroidal rotation and other large-scale phenomena \cite{Camenen_rotation}; Coulomb collisions, which induce neoclassical transport through random scattering; electromagnetic fluctuations, representing turbulence, in the spirit of gyrokinetic theory \cite{Brizard_hahm_gyrokinetic}.

These dynamical components are inherently distinct: the neoclassical drifts are deterministic and large-scale, the collisions are random and uncorrelated, and the turbulent motion is stochastic. Consequently, the particle trajectories are also stochastic (correlated on small scales), thus, the name of the code. T3ST simulates the gyrocenter dynamics of particles within an ensemble of random fields designed to replicate the statistical properties of turbulent fluctuations. While this approach effectively decouples plasma dynamics from field dynamics, it does not limit T3ST to the passive ions (low-concentration impurities). If the synthetic turbulence accurately replicates real turbulence, T3ST can investigate the dynamics of bulk ions as well.

With the characteristic trajectories computed, the transport coefficients are determined as a double average of Lagrangian quantities: over the ensemble of turbulent field realizations (statistical average) and over velocity space (kinetic average).

\subsection{Dynamical equations}
\label{Section_2.1}

We consider a species of charged particles (ions, altough T3ST can be used also for electrons) with mass $m=A m_i$ and charge $q=Z|e|$ that moves within a tokamak electromagnetic environment. $A,Z$ represent mass and ionization numbers, while $m_i, |e|$ are the elementary mass and charge of a hydrogen ion. The strong, macroscopic, nature of the magnetic field $\mathbf{B}$ validates gyrokinetic orderings \cite{Brizard_hahm_gyrokinetic}. In particular, we assume here the high-flow ordering \cite{sugama2017modern,brizard_rotating} with the inclusion of polarization drift effects \cite{wang_hahm_polarization}. Consequently, the real 6D particle trajectory $(\mathbf{x},\mathbf{v})$ is stripped of its fast Larmor rotation and it is described, instead, by 5D gyrocenter coordinates $(\mathbf{X},v_\parallel,\mu)$ (the modern path toward gyrocenter phase-space resorts to Lie's perturbation theory \cite{Littlejohn}). The one body distribution function $f(\mathbf{X},v_\parallel,\mu,t)$ of the ion species is known to obey the gyrokinetic equation \cite{Brizard_hahm_gyrokinetic}

\begin{align}\label{eq_2.1.1}
\partial_t f + \mathbf{v}\nabla_\mathbf{X}f + a_\parallel\partial_{v_\parallel}f = \sum_sC[f,f_s]
\end{align}
where the collisionless equations of motion for gyrocenter coordinates are:

\begin{eqnarray}
	\label{eq_2.1.2a}
\frac{d\mathbf{X}}{dt}&=&\mathbf{v} = v_\parallel\frac{\mathbf{B}^\star}{B_\parallel^\star} + \frac{\mathbf{E}^\star\times\mathbf{b}}{B_\parallel^\star}\\
	\label{eq_2.1.2c}\frac{dv_\parallel}{dt} &=&a_\parallel  = \frac{q}{m}\frac{\mathbf{E}^\star\cdot\mathbf{B}^\star}{B_\parallel^\star}\\
	\label{eq_2.1.2d}\frac{d\mu}{dt} &=& 0
\end{eqnarray} 

and the effective electromagnetic fields, $\mathbf{E}^\star = -\nabla \Phi^\star-\partial_t\mathbf{A}^\star, 
\mathbf{B}^\star = \nabla\times\mathbf{A}^\star$, read:

\begin{eqnarray}\label{eq_2.1.3a}
q\Phi^\star &=&  \frac{m v_\parallel^2}{2} - \frac{m \mathbf{u}^2}{2} + \mu B + q\phi_1^{neo} + q\phi_1^{gc}\\
\label{eq_2.1.3b}\mathbf{A}^\star &=& \mathbf{A}_0 + \frac{m}{q}\left(v_\parallel\mathbf{b}+\mathbf{u}+\mathbf{v}_E\right)+ \mathbf{A}_1^{gc} 
\end{eqnarray} 

The zero order vector potential $\mathbf{A}_0$ is linked to the equilibrium magnetic field via $\mathbf{B}=\nabla \times \mathbf{A}_0$. We consider here only axisymetric plasma equilibria, thus, $\mathbf{B}$ can be expressed in a mixed co-contravariant representation as $ \mathbf{B}=F(\psi)\nabla\varphi + \nabla \varphi\times\nabla\psi$. $\varphi$ is the toroidal angle, part of the right-handed systems of cylindrical $(R,Z,\varphi)$ or toroidal $(r,\theta,\varphi)$ coordinates (COCOS=2 convention \cite{SAUTER2013293}). The poloidal flux function $\psi$ is $\varphi$ invariant due to axisymetry, i.e. $\psi(R,Z)\equiv\psi(r,\theta)$.  Furthermore, we define the \emph{poloidal safety factor} $q_\theta(\psi,\theta)$, the \emph{safety factor} $\bar{q}(\psi)$ and the \emph{generalized poloidal angle} $\chi(\psi,\theta)$ as:

\begin{eqnarray}\label{eq_2.1.4}
q_\theta(\psi,\theta) &=& \frac{\mathbf{B}\cdot\nabla\varphi}{\mathbf{B}\cdot\nabla\theta}\\
\bar{q}(\psi) &=& \frac{1}{2\pi}\int_0^{2\pi}q_\theta(\psi,\theta)~d\theta\\
\frac{\partial\chi}{\partial\theta}\bigg|_\psi &=& \frac{q_\theta(\psi,\theta)}{\bar{q}(\psi)}.
\end{eqnarray} 

With these definitions, the Clebsch representation \cite{Xanthopoulos} of the magnetic field follows: $\mathbf{B}=\nabla (\varphi-\bar{q}\chi)\times\nabla\psi$. For purposes related to a proper representation of small scale turbulent perturbations, most gyrokinetic codes \cite{GORLER20117053,PEETERS20092650,LANTI2020107072} employ the so-called \emph{field-aligned coordinates} $(x,y,z)$ \cite{Beer_fieldaligned}. These are interpreted as radial, "binormal" and parallel coordinates ($\mathbf{B} \propto \nabla y \times \nabla x \propto \partial\mathbf{r}/\partial z$) and are, explicitely:
\begin{eqnarray}\label{eq_2.1.5}
x &=& C_x f(\psi)\\
y &=& C_y (\varphi - \bar{q}\chi)\\
z &=& C_z\chi.
\end{eqnarray}
In practice, T3ST uses the constant values $C_x= a$ (the tokamak minor radius), $C_y = r_0/\bar{q}(r_0)$, ($r_0$ some reference radial position), $C_z=1$ and $f(\psi)  = \rho_t = \sqrt{\Phi_t(\psi)/\Phi_t(\psi_{edge})}$ where $\rho_t\in [0,1]$ is the \emph{normalized effective radius} evaluated with the aid of the toroidal magnetic flux $\Phi_t(\psi) = \iint_{\psi_{axis}}^\psi \mathbf{B}\cdot\mathbf{e}_\varphi dS(\psi^\prime)$ and grossly understood as $\rho_t\approx r/a$. 

Neoclassical theory shows \cite{Hinton_neoclassical} that the zero order MHD equilibrium is both toroidal invariant and constant across magnetic surfaces: the plasma density, the pressure, the ion and electron temperatures, are all functions of $\psi$ (or, equivalently, $\rho_t$, or $x$): $n(\psi), P(\psi), T_i(\psi),T_e(\psi)$. Another zero order effect is the possible existence of a large-scale electric field $\mathbf{E}_0$ that, via the $E\times B$ drift, contributes to the toroidal plasma velocity $\mathbf{u} = R^2\Omega_t\nabla\varphi$. The angular frequency $\Omega_t(\psi)$ can have non-zero shearing components, $d\Omega_t/d\psi\neq 0$. A related, subtle, but very important aspect is that the equations of motion \eqref{eq_2.1.2a}-\eqref{eq_2.1.2c}, are actually expressed in a moving reference frame that rotates as a rigid body with $\mathbf{u}$ \cite{brizard_rotating,Peeters_rotation}.

Up to first order, poloidal flows are known to be strongly damped \cite{10.1063/1.1694506}, thus, we neglect them here. Yet, a neoclassical poloidally dependent electric potential $\phi_1^{neo}(\psi,\theta)$ emerges in toroidally rotating plasma in order to mantain quasineutrality \cite{Hinton_neoclassical}: 
\begin{align}\label{eq_2.1.6}
\phi_1^{neo} = \frac{m\Omega_t(\psi)^2}{2|e|(1+T_i/T_e)}\left(R^2 - \langle R^2\rangle_\theta\right),
\end{align} 
where $\langle \star\rangle_\theta = (2\pi)^{-1}\int_0^{2\pi}\star~d\theta $ denotes poloidal averaging. 

It must be emphasized that, in the absence of collisions and for time-independent perturbations, the energy $E = q\Phi^\star = m v_\parallel^2/2 - m \mathbf{u}^2/2 + \mu B + q\phi_1^{neo}$, the magnetic moment $\mu = mv_\perp^2/2B$ and the canonical toroidal momentum $P_c = \psi - m v_\parallel F(\psi)/qB$ are invariants of motion for the eqns. \eqref{eq_2.1.2a}-\eqref{eq_2.1.2c}.

Coulomb collisions are captured in the collisional operator $C[f,f_s]$ where sumation over all other charged particle species "s" is imposed in eq. \eqref{eq_2.1.1}. The nature of the operator in gyrocenter coordinates follows standard Fokker-Planck formalism $C[f,f_s]  = -\partial_z\left(\mathcal{K}^z f- \mathcal{D}^{zz'}\partial_{z'}f\right)$ with $\mathcal{K}^z,\mathcal{D}^{z,z'}$ to be discussed later \eqref{Section_2.3}.

The central motivation for the T3ST code is the existence of first order terms $\phi_1^{gc},\mathbf{A}_1^{gc}$ in the expressions of $\Phi^\star,\mathbf{A}^\star$ (\eqref{eq_2.1.3a}-\eqref{eq_2.1.3b}). $\phi_1^{gc},\mathbf{A}_1^{gc}$ are electromagnetic potentials that capture electrostatic, respectively magnetic, turbulent fluctuations and, potentially, small external perturbations such as RMPs \cite{Evans2006}. The "\emph{gc}" superscript stands for "gyrocenter" and emphasizes that $\phi_1^{gc}$ fields are evaluated at the level of gyrocenters, thus, require a gyroaverging procedure of the real $\phi_1,\mathbf{A}_1$ to include finite Larmor radius effects. The relation between real and "\emph{gc}" fields is given in terms of their Fourier components, $\tilde{\phi}_1^{gc}(\mathbf{k},t) = \tilde{\phi}_1(\mathbf{k},t)J_0(k_\perp\rho_L)$ where $\rho_L=mv_\perp/qB = \sqrt{2m\mu/q^2B}$ is the particle's Larmor radius, $k_\perp=|\mathbf{k}_\perp|$, $\mathbf{k}_\perp = \mathbf{k} - k_\parallel\mathbf{b}$, $k_\parallel = \mathbf{k}\cdot \mathbf{b}$ and $\mathbf{b}=\mathbf{B}/B$. We note that in practice, $\phi_1$ is designed to capture low-k drift-type turbulence, ITG and TEM \cite{Merz_2010}, since other instabilities (ETG, for example) have a minimal contribution to ion transport. 

\subsection{Test-particle sampling}
\label{Section_2.8}

T3ST is not concerned with solving the GK equation \eqref{eq_2.1.1}, nor with the evaluation of the distribution function $f$ or the electromagnetic fields. Nonetheless, at the heart of the code lies the so-called method of characteristics: the distribution function is conserved along phase-space trajectories of eq. \eqref{eq_2.1.1}. This implies that we can use a test-particle sampling of the distribution function as: 

\begin{eqnarray}
	\label{eq_2.8.1}
	f(z,t) = \sum_{i=1}^{N_p} \frac{1}{J(z)}\delta[z-Z_i(t)]
\end{eqnarray}
where by $z \equiv \left(\mathbf{X},v_\parallel,\mu\right), J(z)=B_\parallel^\star = \mathbf{B}^\star\cdot\mathbf{b}, Z_i(t)$ we denote gyrocenter phase-space coordinates respectively the Jacobian of the gyrocenter transformation and a characteristic trajectory. The expansion \eqref{eq_2.8.1}, in the limit $N_p\to\infty$, represents the test-particle numerical resolution of the distribution function, solution of \eqref{eq_2.1.1}, provided that:
\begin{eqnarray}
	\label{eq_2.8.2}
	\frac{d}{dt}Z(t) \equiv \frac{d}{dt}\left(\mathbf{X}(t),v_\parallel(t),\mu(t)\right) \equiv \left(\mathbf{v}[Z(t)],a_\parallel[Z(t)],0\right).
\end{eqnarray}

We stress, again, that these test-particles are not used to evaluate directly the distribution function or macroscopic quantities, but rather transport coefficients. This is explained in the next section.

\subsection{The transport picture and transport coefficients}
\label{Section_2.7}

The transport component that is directly related to confinement in tokamaks is the "radial" transport, or, more generally, transport across magnetic surfaces. Zero-order neoclassical equilibrium states are Maxwell-Boltzmann distributed, in virtue of H-theorem and the existence of collisions. Supplementary, the radial transport is assumed to be local, on space scales much smaller than neoclassical gradients ($|\nabla \ln n_0|^{-1}$).

For all these reasons, T3ST can implement the following standard scenario for transport investigations: the initial distribution function $f(x,y,z,v_\parallel,\mu)$ is considered Maxwell-Boltzmann $f\sim \sqrt{E/T}exp\left(-E/T\right)$ in the energy space and highly localized along a flux-tube in the physical space ($f\sim \delta(x-x_0) \sim \delta(\psi-\psi_0)$). This distribution is sampled by test-particles whose trajectories are followed in time.

The correct method of defining transport coefficients in non-equilibrium inhomogeneous plasmas both for particles and heat is to be discussed in a separate, future work \cite{palade_2025_inhomogeneous}. Here, we make progress by considering valid the local Fick's like laws for matter fluxes ($\Gamma_n$, from the continuity equations $\partial_t n+\nabla_x\cdot\Gamma_n = 0$): 
\begin{align}
	\label{eq_2.7.1}
	\Gamma_n(t|x)  &= V_n(t|x)n(x) - D_n(t|x)\partial_x n(x)
	\end{align}

We define $X(t|x_0) = \nabla x\cdot \mathbf{X}(t), X(0|x_0) = x_0$, essentially, the "radial" ($x$) coordinate of a test-particle at the time $t$ that started from $x_0$ at $t=0$. With these, the radial transport coefficients are defined as ensemble $\{\phi_1\}$, velocity-space $\{v_\parallel,\mu\}$ and $\{y,z\}$ Lagrangian averages $\langle\rangle$:
\begin{align}
	\label{eq_2.7.2}
	V_n(t|x) &= \frac{d}{dt}\langle X(t|x)\rangle\\
	\label{eq_2.7.3}
	D_n(t|x) &= \frac{1}{2}\frac{d}{dt}\left(\langle X(t|x)^2\rangle-\langle X(t|x)\rangle^2\right)
	\end{align}

We identify the particle pinch $V_n$, diffusion $D_n$
. These quantities tend to sature on microscopic time-scales in most regimes of interest, thus, only their asymptotic values $V(x) = \lim\limits_{t\to\infty}V(t|x),D(x) = \lim\limits_{t\to\infty}D(t|x)$ are relevant for the characterization of transport.

\subsection{Magnetic configurations}
\label{Section_2.2}

The general representation of equilibrium axisymmetric magnetic field is $\mathbf{B}=F(\psi)\nabla\varphi +\nabla\varphi\times\nabla\psi$. However, it is now necessary to describe particular magnetic models (specify $\psi$ and $F(\psi)$) that T3ST can implement, namely: circular, Solov'ev and experimentally reconstructed.

\subsubsection{The circular model}
\label{Section_2.2.1}

The circular model is the simplest, analytical, magnetic configuration possible and, for that, it has been used previously in many investigations. It is not an exact solution of the Grad-Shafranov equation \cite{shafranov1958magnetohydrodynamical,Atanasiu}, but rather an approximation valid in the core of plasmas or in the low-aspect-ratio limit $\varepsilon=r/R_0\ll 1$. The covariant representation of the magnetic field is
\begin{align}\label{eq_2.2.1}
\mathbf{B} = B_0 R_0\left(\nabla\varphi + \frac{r b_\theta(r)}{R}\nabla\theta\right)
\end{align} 
where $B_0$ is the magnitude of $\mathbf{B}$ at the magnetic axis ($Z=0,R=R_0$) and $b_\theta(r)$ is a measure of the poloidal component. One can easily identify that this form \eqref{eq_2.2.1} corresponds to $F(\psi) = B_0 R_0$ and  
\begin{align}
	\label{eq_2.2.2a}
\psi(r) &= B_0 R_0 \int_0^r b_\theta(r^\prime) ~dr^\prime\\
\label{eq_2.2.2b}b_\theta(r) &= \frac{r}{\bar{q}(r)\sqrt{R_0^2-r^2}}\\ 
\label{eq_2.2.2c}\chi(r,\theta) &= 2\arctan\left(\sqrt{\frac{R_0-r}{R_0+r}}\tan\left(\frac{\theta}{2}\right)\right).
\end{align} 

In practice, analytical expression of $\bar{q}$ (such as $\bar{q}(r) = c_1 + c_2 r + c_3 r^2$) are used. The radial field aligned coordinate reads:

$$x = C_x \rho(\psi) = a \sqrt{\frac{1-\sqrt{1-\frac{r^2}{R_0^2}}}{1-\sqrt{1-\frac{a^2}{R_0^2}}}} \approx r .$$

\subsubsection{Solov'ev equilibria}
\label{Section_2.2.2}

These are a class of analytical, global, solutions of the Grad-Shafranov equation obtained under mild assumptions regarding the linearity of pressure $\mu_0dP/d\psi = A$ and current $dF^2(\psi)/d\psi = 2AR_0^2\gamma/(1+\alpha^2)$ profiles. The expression of the poloidal flux function in cylindrical coordinates $(R,Z)$ reads \cite{solov1968theory}:
\begin{eqnarray}\label{eq_2.2.3}
\psi = \frac{A}{2(1+\alpha^2)}\left(\left(R^2-\gamma R_0^2\right)Z^2+\frac{\alpha^2}{4}\left(R^2-R_0^2\right)^2\right).
\end{eqnarray} 

The parameter $R_0$ is the radial position of the magnetic axis, the $\alpha,\gamma$ parameters are related to the minor radius $a=R_0/2\left(\sqrt{2-\gamma}-\sqrt{\gamma}\right)$ and the elongation $\kappa = \alpha /\left(1-\sqrt{\gamma/(2-\gamma)}\right)$ of the plasma, while the current function is
\begin{align}\label{eq_2.2.4}
	F(\psi) = B_0 R_0 \sqrt{1+\psi\frac{2 A\gamma}{B_0^2(1+\alpha^2)}}.
\end{align}

Unfortunatelly, there are no available analytical expression for the generalized poloidal angle $\chi$ or the normalized radius $\rho_t(\psi)$. Consequently, whenever T3ST implements Solovev equilibrium, it uses $\rho_t = \sqrt{\psi/\psi(a)}$ and $\chi (r\to x)$ given by eqns. \eqref{eq_2.2.2a}-\eqref{eq_2.2.2c}. This approximation reflects solely (and mildly) on the evaluation of turbulent fields since the neoclassical drifts are evaluated from cylindrical, not field-aligned, coordinates.

\subsubsection{Experimentally reconstructed equilibria}
\label{Section_2.2.3}

By \emph{experimentally reconstructed equilbria} we refer to magnetic data used by the tokamak community in convetional format files such as G-EQDSK. The latter contains the poloidal flux function $\psi$ provided numerically on a $(R,Z)$ grid and $F(\psi),\bar{q}(\psi)$ on a $\psi$ grid. These quantities are solutions of the Grad-Shafranov equation complemented by experimental measurements (current and pressure profiles, boundary conditions, etc.). There are multiple codes available for such equilibrium reconstruction, CHEASE \cite{LUTJENS1996219}, EFIT \cite{Lao_1985}, PLEQUE \cite{kripner2019towards}, NICE \cite{FAUGERAS2020112020}, EQUINOX \cite{BLUM2012960}, etc. Thus, whenever T3ST is used to investigate accurately specific tokamak discharges, this is the choice of preference for the magnetic equilibrium.

\subsection{Collisions}
\label{Section_2.3}

The Fokker-Plank operators for Coulomb collisions, $C[f,f_s]  = -\partial_z\left(\mathcal{K}^z f- \mathcal{D}^{zz'}\partial_{z'}f\right)$, have been worked out previously in literature for gyrocenter coordinates \cite{Brizard_collision}. Since T3ST follows particle trajectories $\{\mathbf{X}(t),v_\parallel(t),\mu(t)\}$, we are obliged to describe the effects of collisions at their level. This is known as Monte-Carlo representation of the collisional operator \cite{Hirvijoki_monte_carlo_coll} and has been implemented in other particle codes such as ASCOT \cite{HIRVIJOKI20141310}. We state here only that collisions change the nature of gyro-center dynamics, from ODE to stochastic SDE:
\begin{eqnarray*}\label{eq_2.3.1}
d\mathbf{X} &=& \sqrt{2D_c} \left(\mathbf{I}-\mathbf{b}\otimes\mathbf{b}\right)d\mathcal{W}^{\mathbf{X}}\\ dv_\parallel &=&  v_\parallel\left(-\nu + \left(2\frac{D_\parallel - D_\perp}{v^2}+\frac{\partial D_\parallel}{v\partial v}\right)\right) dt +\\ &+&\Sigma^{v_\parallel,v_\parallel}d\mathcal{W}^{v_\parallel}+ \Sigma^{v_\parallel,\mu}d\mathcal{W}^{\mu} \\
d\mu &=& \mu\left( -2\nu  + \frac{m}{\mathcal{E}}\left(v_\parallel\frac{\partial D_\parallel}{\partial v}+ \frac{3(D_\parallel - D_\perp)\mu B+ 2\mathcal{E}D_\perp}{\mu B}\right)\right) dt +\\ &+& \Sigma^{\mu,v_\parallel}d\mathcal{W}^{v_\parallel}+ \Sigma^{\mu,\mu}d\mathcal{W}^{\mu} 
\end{eqnarray*} 
where $d\mathcal{W}^i$ are differential independent stochastic Wiener processes with zero mean. The explicit expressions of all other quantities involved are detailed in \eqref{A1}. We also note that $\left(E=q\Phi^\star,P_c,\mu\right)$ are no longer invariants of motion under collisions. Consequently, the time evolution of $\mu$ must also be integrated numerically. 

Additionally, T3ST can implement the much simpler Lorentz collisional operator which is best suited for electron transport but is a good estimate also for ions in coronal equilibrium with the plasma. This choice corresponds to the stochastic evolution of the pitch angle $\lambda = v_\parallel/v, v=\sqrt{v_\parallel^2+v_\perp^2}$ and kinetic energy $E_{kin}$ \cite{boozer1980monte}:
\begin{align}\label{eq_2.3.2}
	\lambda (t+\Delta t) &= \lambda (t) - \nu_d\lambda (t)\Delta t +\sigma \sqrt{(1-\lambda^2 (t))\nu\Delta t}\\
	E_{kin} (t+\Delta t) &= E_{kin} (t) - 2\nu_e\Delta t E_{kin} (t)T\left(\frac{1}{T}-\frac{3}{2E_{kin}} - \frac{d \ln\nu_e}{dE_{kin}}\right) +2\sigma \sqrt{\nu_e\Delta tE_{kin} (t)T}
\end{align} 
where $\sigma = \pm 1$, randomly chosen for each particle at each time-step. The equations \eqref{eq_2.3.2} are designed to preserve homogeneous pitch angle and Maxwell-Boltzmann energetic distributions, i.e. $P(\lambda,t+\Delta t) = P(\lambda)$.

\subsection{(Turbulent) perturbations}
\label{Section_2.4}

The central motivation for developing T3ST is the characterization of turbulent transport. The latter is driven by turbulent fields, represented in the equations of motion (\eqref{eq_2.1.2a}-\eqref{eq_2.1.2c}), indirectly, via the first-order perturbative electromagnetic potentials $\phi_1^{gc}, \mathbf{A}_1^{gc}$. $\phi_1^{gc}$ is a scalar potential reserved for electrostatic fluctuation, in particular low-k drift type turbulence, ITG and TEM \cite{Merz_2010}, while other high-k instability originating turbulence (such as ETG \cite{PhysRevLett.85.5579}) are ignored since they are not relevant for ion dynamics. $\mathbf{A}_1^{gc}$, a vector field, is reserved for magnetic fluctuations which can be of turbulent nature or external perturbations such as RMPs \cite{Evans_2015}. 

The following discussion focuses on the case of electrostatic turbulence. Nonetheless, most of the details are  transferable to the magnetic case.

\subsubsection{Electrostatic turbulence}
\label{Section_2.4.1}

T3ST does not aim at computing self-consistently the perturbed fields from plasma dynamics, but rather model them statistically as an ensemble of random fields $\{\phi_1,\mathbf{A}_1\}$ (not "gc") with given statistical properties. The latter are, especially for ITG/TEM, quite consistent across multiple plasma experiments. In order to justify the technical representation of the random fields, we have to make a few benign assumptions.

First, we assume that the statistics of real field fluctuations is normal, i.e. the PDF is Gaussian $P[\phi_1]\sim \exp\left(-\phi_1^2/2\langle\phi_1^2\rangle\right)$. The same holds true for derivatives $\partial_x\phi_1$.  While most gyrokinetic, fluid or experimental investigations suggest that turbulent fluctuations in tokamak devices exhibit a small departure from normal distribution, the impact of skewness or kurtosis on transport has been investigated elsewhere \cite{palade_pom_2022} and found to be minimal.

The second assumption, motivated by the microscopic nature of fluctuations, is that the turbulence is homogeneous. Together with Gaussianity, this implies that the autocorrelation $\mathcal{E}(\mathbf{r},t\big| \mathbf{r}^\prime,t^\prime) = \langle \phi(\mathbf{r},t)\phi(\mathbf{r}^\prime,t^\prime)\rangle = \mathcal{E}(\mathbf{r} - \mathbf{r}^\prime\big| t-t^\prime)$ is the only piece of statistical information needed and can be evaluated as Fourier transform of the power spectrum $S(\mathbf{k},\omega) = \langle \big|\tilde{\phi_1}(\mathbf{k},\omega)\big|^2\rangle$. 

The last hypothesis is related to the fact that fluctuations of interest (ITG and TEM) originate from drift-type instabilities which have linear dispersion relations $\omega_\star(\mathbf{k})$ that hold approximately true also in the saturated regime. In the current version of the code we use a simple, slab-like formula ($\mathbf{V}^\star_s=-\nabla p_s\times \mathbf{b}/(n_s q_s B), \rho_s$ are the "s"(ion-ITG or electron-TEM) diamagnetic velocity, respectively the "s" Larmor radius): 

\begin{align}
	\label{eq_2.4.3}
	\omega_\star(\mathbf{k}) = \frac{\mathbf{k}\cdot\mathbf{V}^\star_s}{1+\rho_s^2|k_\perp|^2}
\end{align}

Given all these (mild) assumptions, T3ST constructs a statistical ensemble of random fields $\{\phi_1(\mathbf{x},t)\}$ from an associated ensemble of Fourier-space white noises $\{\eta(\mathbf{k},\omega)\}$, where the statistical averages read  $\langle \eta(\mathbf{k},\omega)\eta(\mathbf{k}^\prime,\omega^\prime)\rangle = \delta(\mathbf{k}^\prime+\mathbf{k})\delta(\omega+\omega^\prime)$. The recipe is to define the Fourier components as $\tilde{\phi}_1(\mathbf{k}) = \sqrt{S(\mathbf{k},\omega)}\eta(\mathbf{k},\omega)$. One can easily show \cite{Palade2021} that the resulting fields are random, Gaussian and reconstruct the correct correlation function $\mathcal{E}(\mathbf{x}-\mathbf{x}^\prime,t-t^\prime)$. 

In principle, random fields are best represented in toroidal coordinates $(r,\theta,\varphi)$ through a double-Fourier series that captures clearly the required periodicity of any real field:
\begin{align}
	\label{eq_2.4.0}
	\phi(\mathbf{x},t) = \sum_n \sum_m \phi_{n,m}(r,t)e^{i\left(n\varphi+m\theta\right)}.
\end{align}

On the other hand, turbulence in tokamaks is characterized by long correlation (wavelengths) along the field-lines and short correlations in the perpendicular plane. This could enable one to use the so called ballooning representation \cite{PhysRevLett.40.396}. Nonetheless, in the world of gyrokinetics (local or global) the method of preference is the use of field-aligned coordinates \cite{Beer_fieldaligned}. Since for T3ST the properties of turbulence are inspired by experimental data and gyrokinetic results, we choose for consistency to evaluate $\phi_1$ also in field-aligned coordinates, even though all magnetic drifts are computed in cylindrical coordinates.

The formal representation of random fields used by T3ST is the following:
\begin{align}\label{eq_2.4.1}
	\phi_1(x,y,z,t) &= \int~d\mathbf{k} ~d\omega\tilde{\phi}_1(\mathbf{k}, \omega)e^{i\left(k_x x+k_y y+k_z z-(\omega_\star(\mathbf{k}) +\omega) t\right)}
\end{align}
where $y=C_y(\varphi-\bar{q}(x)\chi)$. Given that $k_\parallel\ll k_\perp$, the following connection with the toroidal representation is approximately true: $k_y \approx n/C_y, m = -[\bar{q}(x_0)n]+\Delta m,  k_z = (\Delta m+\{\bar{q}(x_0)n\})/C_z$, where $[], \{\}$  denote the integer part respectively the fractional part and $n,\Delta m\in\mathbb{Z}$. 

Regarding the explicit forms of the spectrum $S(\mathbf{k},\omega)$, they are essentially saturation rules \cite{10.1063/1.4954905,Dudding_2022}. We use an analytical expression designed to capture the gross feature of drift-type ITG/TEM turbulence that stem from the shape of growing rates:
\begin{eqnarray}\label{eq_2.4.2}
S &=& A_\phi^2\frac{\tau_c\lambda_x\lambda_y\lambda_z}{\left(2\pi\right)^{5/2}} \frac{e^{-\frac{k_x^2\lambda_x^2+k_z^2\lambda_z^2}{2}}}{1+\tau_c^2\omega^2}\frac{k_y}{k_0}\left(e^{-\frac{(k_y-k_0)^2\lambda_y^2}{2}}-e^{-\frac{(k_y+k_0)^2\lambda_y^2}{2}}\right)
\end{eqnarray}
where $\lambda_x,\lambda_y,\lambda_z$ are correlation lengths along the field-aligned coordinates, $k_0$ is a control wavenumber that, together with $\lambda_y$, refines the position of the most-unstable mode, $k_{y}^{max}\approx k_0/2+\sqrt{k_0^2/4+2/\lambda_y^2}$. The parameter $\tau_c$ is a correlation time that measures departure of real frequencies of modes from the dispersion relation while $A_\phi$ is the turbulence strenght. Note that $S/A_\phi^2$ is normalized to $1$, since $\mathcal{E}(\mathbf{0},0)=\langle\phi_1^2(\mathbf{0},0)\rangle = A_\phi^2$. Typical values of the parameters are $\lambda_x\sim 5\rho_i, k_0\sim 0.1\rho_i^{-1}, \lambda_y\sim 5 \rho_i,\lambda_z\sim 1,\tau_c\sim R_0/v_{th}, |e| A_\phi\sim 1\% T_i$. Note that $\lambda_z\sim 1$ implies $\lambda_\parallel \sim \bar{q}R_0$ in line with experimental evidence \cite{10.1063/1.859070}.

Finally, in general scenarios, the drift turbulence in tokamak plasmas is actually a super-position of ITG and TEM, in distinct fractions $A_i$, respectively $A_e$ \cite{Palade_2023,Merz_2010}. For this reason, T3ST is endowed with the ability to generate total electrostatic fields as \cite{Palade_2023}:

$$\phi_1 = \sqrt{A_i}\phi_1^{ITG}+\sqrt{A_e}\phi_1^{TEM}$$

with $A_i+A_e=1$.


\subsection{Particle initial conditions}
\label{Section_2.6}

T3ST allows, in principle, for the initialization of particle markers (equivalent to the particle distribution function $f(\mathbf{X},v_\parallel,\mu,t=0)$) to any configuration corresponding to any physical scenario envisaged. Yet, for the purpose of turbulent transport investigations and for consistency with different assumptions of the code, there are currently only a handfull of possibilities implemented. Before detailing them, let's recall that the gyrocenter energetic coordinates $(v_\parallel,\mu)$ are equivalent to the kinetic energy-pitch angle coordinates $(E_k,\lambda)$ given that $\mu=m \mathbf{v}_\perp^2/2B, E_k=mv^2/2, \mathbf{v}=\mathbf{v}_\perp + \mathbf{b}v_\parallel$ and $\lambda = v_\parallel/v$.

We use the factorization $f(\mathbf{X},v_\parallel,\mu,t=0) = n(\mathbf{X})F(v_\parallel,\mu|\mathbf{X})$ where $n(\mathbf{X})$ is test-particle density and the energetic distribution $F(v_\parallel,\mu|\mathbf{X})\equiv F(E_k|\mathbf{X})g(\lambda)$ is normalized $\iint dv_\parallel d\mu JF(v_\parallel,\mu|\mathbf{X}) = 1$ with $J=B_\parallel^\star$ the Jacobian of the gyrocenter transformation.

\subsubsection{Initial space distributions}
\label{Section_2.6.1}

\begin{itemize}
	\item Point-distributions, characterized by $n(\mathbf{X})=\delta(\mathbf{X}(t=0)-\mathbf{X}_0)$ when all particles are placed at a single space point $\mathbf{X}_0$. This choice can be a usefull for the evaluation of dynamical Green-functions $\langle h(\mathbf{X}(t_1),t_1;\mathbf{X}(t_2),t_2;...)\rangle$. \item Flux surface distributions, characterized by $n(\mathbf{X}) = \delta(\psi(\mathbf{X}(t=0))-\psi_0)$ when all particles are spread uniformly across a flux surface of value $\psi_0$. Mathematically, such geometrical locus is the solution of $\psi(R,Z) =\psi(r,\theta) = \psi_0$ or $x=x_0(\psi_0)$. This is the choice of preference for most transport simulations.
\end{itemize}

\subsubsection{Initial kinetic energetic distribution}
\label{Section_2.6.2}

\begin{itemize}
	\item Point distributions, characterized by $F(E_k|\mathbf{X}) = \delta(E_{kin}-E_0)$ when all particles, regardless of their space position, have the same kinetic energy. This choice suits investigations of transport over the energetic space. 
	\item Maxwell-Boltzmann distribution, i.e.  $F(E_k|\mathbf{X})  = \sqrt{E_{kin}/T}exp(-E_{kin}/T(\mathbf{X}))$. This requires specifying the species's temperature $T(\mathbf{X})$. This is the choice of preference for most transport studies, sometimes together with the assumption of coronal equilibria, i.e. $T=T_i$.
\end{itemize}

\subsubsection{Pitch angle distribution}
\label{Section_2.6.3}

\begin{itemize}
	\item Point distributions, characterized by $g(\lambda) = \delta(\lambda-\lambda_0)$ when all particles have the same pitch angle $\lambda_0$. Coupled with point distributions of energy, it allows for the investigation of a single neoclassical trajectory over multiple turbulent field realizations.
	\item Uniform distributions, characterized by $g(\lambda) \sim \Theta(1-|\lambda|)$ when the pitch angle is distributed uniformly in the $[-1,1]$ domain. This implies an isotropic distribution in the $(\mathbf{v}_\perp,v_\parallel)$ space and it is used mostly in conjunction with a Maxwell-Boltzmann distribution of energies.
\end{itemize}

Once the pitch angles and kinetic energies are generated, together with the initial positions of particles, one can easily evaluate $B$ and the gyro-center coordinates $v_\parallel,\mu$.

\subsection{Probabilities of stochastic trajectories}
\label{Section_2.5}

 Up to this point, we have described how T3ST is equipped with the ability to use various initial energetic distributions for the kinetic energy $E_{kin}$. However, $E_{kin}=m v_\parallel^2/2+\mu B$ represents only a fraction of the total energy, given by $q\Phi^\star = E_{kin} - m \mathbf{u}^2/2 + q\phi_1^{neo} + q\phi_1^{gc}$. Since we typically consider equilibrium initial states, the potential Boltzmann distribution should reflect $q\Phi^\star$ rather than $E_{kin}$.
 
 To address this issue, T3ST follows the steps outlined below. First, the initial distribution of particles in physical space, $n(\mathbf{X})$, is determined. Next, the kinetic energies and pitch angles, $F(E_{kin})$ and $g(\lambda)$, are set. At this point, the code is ready to evaluate the magnetic field $B(\mathbf{X})$, the neoclassical field $\phi_1^{neo}(\mathbf{X})$, and compute $v_\parallel$ and $\mu$. The random fields $\phi_1^{gc}$ are handled separately, as they are normally distributed and independent of the particles. Their values are evaluated at the positions of their corresponding trajectories, $\phi_1^{gc}(\mathbf{X},t=0)$. 
 
 For each particle and field realization, a probabilistic weight $P = \exp\left(-q(\phi_1^{gc}+\phi_1^{neo})/T\right)$ is assigned. Finally, all weights are normalized to unity, $P = P/\langle P\rangle$.

\section{Numerical details}
\label{Section_3}

\subsection{Coordinate systems representations}
\label{Section_3.1}

Within the nuclear fusion community there are multiple choices of coordinate systems. For example, one can describe the geometrical setup via toroidal $(r,\theta,\varphi)$, cylindrical $(R,Z,\varphi)$ or field-aligned \cite{Beer_fieldaligned} $(x,y,z)$ coordinates with different sign and scaling conventions \cite{SAUTER2013293}. 

When solving the equations of motion \eqref{eq_2.1.2a}-\eqref{eq_2.1.2c} for gyrocenters $\{\mathbf{X},v_\parallel\}$, T3ST represents the space position $\mathbf{X}$ in the right-handed, orthogonal, cylindrical coordinate system $\mathbf{X}=(Q_1,Q_2,Q_3)=(R,Z,\varphi)$ with associated Lame coefficients $(h_1,h_2,h_3)= (1,1,R)$. This choice is motivated by the fact that, many times, the magnetic equilibrium is provided by G-EDQSK files which employ a cylindrical grid. The equations of motion \eqref{eq_2.1.2a}-\eqref{eq_2.1.2c} are projected as $dQ_i(t)/dt = \mathbf{v}\cdot\nabla Q_i$, and the effective fields $\mathbf{B}^\star,\mathbf{E}^\star$ \eqref{eq_2.1.3a}-\eqref{eq_2.1.3b} are represented via their contra/co-variant components, $\mathbf{B}^\star = B_\star^k\partial\mathbf{r}/\partial Q_k$, $\mathbf{E}^\star = E_k^\star\nabla Q_k$. When computing the $E\times B$ part of the drifts $\mathbf{v}\cdot\nabla Q_i \sim \left(\mathbf{E}^\star\times\mathbf{b}\right)\cdot\nabla Q_i$, matrix elements of the following type are also needed: 

$$f_{i,j} = (\nabla Q_j\times\mathbf{b})\cdot\nabla Q_i = \varepsilon_{j,k,i}\frac{h_k B^k}{h_i h_j B}.$$

The turbulent fields are represented in (pseudo)-field-aligned coordinates chosen as $x = C_x \rho_t(\psi), y^\prime = C_y (\varphi - \bar{q}(r_0)\chi), z = C_z\chi$ where $C_x= a$, $C_y = r_0/\bar{q}(r_0)$, $C_z=1$. The evaluation of drifts together with their placement within the eoms, requires the computation of two sets of matrix components:  $g_{i,j} = \nabla x_i\cdot \partial\mathbf{r}/\partial Q_j = \partial x_i/\partial Q_j$ for the ExB drift and $m_{i,j} = \left(\nabla x_i\times\mathbf{b}\right)\cdot \partial\mathbf{r}/\partial Q_j$ for the polarization drift.

We underline that $\mathbf{k} = \nabla (\mathbf{k}\cdot\mathbf{r}) = k_i \nabla x_i$. Consequently, $k_\parallel = \mathbf{k}\cdot\mathbf{b} = k_i \nabla x_i\cdot \mathbf{b} = k_i B^j/B \nabla x_i\cdot \partial\mathbf{r}/\partial Q_j = k_i B^j/B g_{i,j}$ while the square of the perpendicular component $k_\perp^2 = k_i k_j \left(\nabla x_i\cdot\nabla x_j\right) - k_\parallel^2$. Given that the diamagnetic velocity $\mathbf{V}^\star \sim \nabla p(\psi)$, it follows that: $\mathbf{V}^\star\cdot\mathbf{k} \propto (\ln  p(x_1))^\prime k_i \nabla x_1\cdot\nabla x_i$.

\subsection{Scaling}
\label{Section_3.2}

In T3ST, the equations of the model \eqref{Section_2} are scaled as follows: space coordinates $(Q_1,Q_2,Q_3)=(R,Z,\varphi)\to (R_0,R_0,1)$, velocities $(v_\parallel,\mathbf{u},\mu)\to (v_{th},v_{th},m_i  v_{th}^2/B_0)$, energies $(E,q\Phi^\star)\to T_i$, field-aligned coordinates $(x,y,z)\to (\rho_i,\rho_i,1)$, wavenumbers $(k_x,k_y,k_z)\to (\rho_i^{-1},\rho_{i}^{-1},1)$, time $t\to R_0/v_{th}$, frequencies $(\omega,\nu,\Omega_t)\to v_{th}/R_0$, magnetic field $B\to B_0$, electric field $E\to m_H v_{th}^2/|e|R_0$, poloidal flux function $\psi\to B_0 R_0^2$, current function $F\to B_0 R_0$, electrostatic potential $\phi_1\to A_\phi$, gradients $\nabla \to R_0^{-1}$. The following definitions have been used: $R_0$ is the radial position of the magnetic axis, $B_0$ is the magnetic field's magnitude at $R_0$, $v_{th} = \sqrt{T_i/m_i}$ the thermal velocity of a $H$ ion at plasma temperature $T_i$ and $\rho_i = m_i v_{th}/|e|B_0$ the Larmor radius of the same ion. Note that $m_i$ is approximately the mass of a proton, even if we investigate a deuterium plasma. For that reason, special care should be taken when expressing results in scaled values.

\subsection{Trajectory propagation}
\label{Section_3.3}

For the numerical solution of the equations of motion (\eqref{eq_2.1.2a}-\eqref{eq_2.1.2c}), T3ST employs a 4th-order Runge-Kutta method. For the collisional component, which transforms the nature of the equations of motion from ODEs to SDEs, we use a direct Euler method corresponding to the Itô interpretation \cite{1944519}. The time domain is discretized over an interval $(0, t_{max})$ into $N_t$ equidistant points.

For most simulations, $t_{max} = 200$, corresponding to a real simulation time of $200 R_0/v_{th} \sim 10^{-3}s$, is sufficient to capture the majority of neoclassical and turbulent dynamics while allowing for adequate relaxation of transport toward asymptotic behavior. In the absence of turbulence, $N_t \approx 10^3$ is adequate for accurate numerical solutions. However, when turbulence is present, the required number of temporal points depends on the magnitude of $A_\phi$. In such cases, up to $N_t = 10^4$ may be needed for $t_{max} = 200$.

\subsection{Numerical resolution}
\label{Section_3.4}

There are four integers that control the accuracy of a T3ST simulation: $N_t,N_p,N_c,N_{real}$. $N_t$ together with the simulation time $t_{max}$ defines the equidistant time-step $\Delta t$ for the integration of the equations of motion, thus, $\Delta t\ll 1$ is vital for accurate, individual, trajectories. 

$N_p$ denotes the number of particles considered in the sampling of the distribution function $f(z,t)\equiv \sum_i^{N_p}J^{-1}(z)\delta(z-z_i(t))$. This numerical parameter is important for an accurate sampling, thus, for the convergence of phase-space averages. Practical experience indicates that $N_p\sim 10^{4-5}/N_{real}$ suffices for particle spreading $\langle X^2\rangle$, but, for an accurate representation of average displacement $\langle X\rangle$, $N_p\sim 10^{5-6}/N_{real}$ particles are required due to the similar magnitudes of $\langle X\rangle$ and numerical fluctuations. 

T3ST employs a statistical description of turbulence via an ensemble of random fields. In practice, the dimension of this ensemble is $N_{real}$. This means that the same $N_p$ particles are propagated in $N_{real}$ realizations of the turbulent fields. This parameter is important for the convergence of statistical averages over the ensemble of random fields. Note that for localized initial distributions $N_{real}$ must be high ($\sim 10^3$) but for the standard case of flux-tube distributed gyrocenters, $N_{real} = 1$ is a perfectly valid numerical choice in T3ST that may give results close to simulations with much higher values of $N_{real}$. 

Finally, $N_c$ stands for the number of partial waves used in the representation of a single random field (see next section \eqref{Section_3.5}). As discussed elsewhere \cite{palade2020fast}, it does not affect the convergence of the ensemble of random fields, nor its effective correlation. It can, however, affect the Gaussianity of the distribution of fields, $P[\phi(\mathbf{r})]$, and, consequently, induce high-order long-range correlations which can be detrimental since they enhance artificially the transport. In general, $N_c\sim 10^2$ is used in simulations.

We note, without proof, that numerical fluctuations (the convergence) of the transport coefficients decays (improves) as $\sim (N_p N_{real})^{-1/2}$. This is rather unfortunate, as the complexity of the code scales linearly as $\mathcal{O}(N_p N_{real})$, thus, a one-hundred-fold in the numerical effort results in only a ten-fold improvement in the convergence. Depending on the scenario considered and on the computational limitations, one might try to balance $N_p$ and $N_{real}$. In practice we routinely use $N_p=10^4$ and $N_{real}=10^{1-2}$. 

The pathological behavior of slow numerical convergence can be partially cured at the level of transport coefficients by using computational techniques that cancel out numerical fluctuations. One method is to average the quantities on a larger time-frame asymptotically $D(x) = T^{-1}\int_t^{t+T} d\tau D(\tau|x)$ or to use the following definitions:
\begin{align}
	V^\prime(t|x) = \frac{\langle X(t)\rangle}{t}\\
	D^\prime(t|x) = \frac{\langle X^2(t)\rangle - \langle X(t)\rangle^2}{2t} .
\end{align} 

\subsection{Turbulent field generation}
\label{Section_3.5}

Turbulent random fields are assumed Gaussian and can be represented in a Fourier decomposition using a white noise \cite{palade2020fast}. In practice, such integrals are discretized in the spirit of a Riemann-sum integration, but with a smart trick that avoids unnecesary numerical efforts and spurious long-range correlations or periodicities. For each field realization we generate $N_c$ pairs of \emph{random} wavenumbers and frequencies $\{\mathbf{k}_i,\omega_i\}$ from a a PDF that is precisely the normalized turbulence spectrum $S(\mathbf{k},\omega)$. The white noise is sampled by random phases $\alpha_i\in[0,2\pi)$. This allows us to evaluate the fields (and their derivatives) as:
\begin{align}\label{eq_3.5.1}
	\partial_x^{(n)}\phi_1^{gc}(\mathbf{X},t) = \sqrt{\frac{2}{N_c}}\sum_i^{N_c}J_0(k_i^\perp\rho_L(\mathbf{X}))(k_i^x)^n\sin\left(\mathbf{k}_i\cdot\mathbf{X}-\omega_i t+\alpha_i+n\pi/2\right)
	\end{align}
where $\sqrt{2/N_c}$ is a normalization factor, $J_0(k_i^\perp\rho_L(\mathbf{X}))$ stems from the Larmor averaging of the fields, $k_i^x$ is the contravariant $x$ component of the wavenumber $\mathbf{k}_i$. 

The Central Limit Theorem ensures us that, in the limit $N_c\to\infty$, the above representation \eqref{eq_3.5.1} is a Gaussian random field. It also has zero mean $\langle\partial_j\phi_1^{gc}(\mathbf{X},t)\rangle = 0$ due to the uniformly random nature of phases $\alpha_i$ and reproduces the correct spectrum, $S(\mathbf{k},\omega)$. For more details, see \cite{palade2020fast}. The effects of non-Gaussianity have been investigated elsewhere \cite{palade_pom_2022} and found to be minimal.

\subsection{Random generation, interpolation and parallelization}
\label{Section_3.6}

T3ST relies fundamentally on a statistical description of turbulence, a kinetic description of the plasma particles and a Monte-Carlo representation of collisions. All these elements, together with numerical necesities and a desire to eliminate any spurious biases, require the generation of random quantities inside the code. 

Regarding the particle initialization, which is essentially the sampling of the distribution function $f(z,t=0)$, we have already mentioned in Section \eqref{Section_2.6} that many possible configurations of particles are in fact distributions of gyrocenters $\{R_i(t=0),Z_i(t=0),\varphi_i(t=0)\}_{i=1,N_p}$, pitch angles $\{\lambda_i(t=0)\}_{i=1,N_p}$ and energies $\{E_i^(kin)\}_{i=1,N_p}$. All these are randomly generated.

Secondly, the evaluation turbulent fields $\phi_1$ requires the generation of random numbers with a given PDF \eqref{Section_3.5}. In practice, given the possibly complicated shapes of the spectrum, T3ST uses acceptance-rejection alogorithms. 

Finally, random generation is also used for collisions. As seen in \eqref{Section_2.3} the Monte-Carlo description of Coulomb collisional operators on individual trajectories resorts to stochastic Ito equations and employ Wiener processes $\mathcal{W}(t_{n+1})$. Their differential forms are, essentially, Gaussian random numbers $\zeta(t_n)$ as $d\mathcal{W}(t_n) = \zeta(t_n)/\sqrt{\Delta t}$.

One of the possible choices for magnetic equilibria is the \emph{experimentally reconstructed equilibria} that uses G-EDSQK files. The latter provide flux functions $\psi(R,Z)$ numerically on an equidistant cartesian $(R,Z)$ grid. To evaluate neoclassical (magnetic) components of motion (\eqref{eq_2.1.2a}-\eqref{eq_2.1.2c}) one has to evaluate such functions on Lagrangian trajectories $R(t),Z(t)$. In order to do that, we use a simple technique of bilinear interpolation via nearest neighbours. 

Practical experience has shown that $\sim 30\%$ of the computing time required for a T3ST simulation is needed for the evaluation of turbulent field's derivatives at particle positions. This can be easily understood given that the numerical complexity scales as $\mathcal{O}(N_{real}\times N_p\times N_t)$ for neoclassical components of motion and as $\mathcal{O}(N_{real}\times N_p\times N_t\times N_c)$ for perturbations. The only method to minimize this considerable numerical effort is to enhance $N_{real}$ for a good memory management and use parallelization techniques.

\section{Testing and results}
\label{Section_4}

The transport model implemented in T3ST involves a comprehensive set of input parameters that can be of numerical nature, related to the model, plasma equilibrium, particle distribution or turbulence. 

In order to test the code, we define the \emph{baseline scenario} as an ITG-dominated tokamak equilibrium corresponding to a typical WEST discharges ($\#54178$). Most of the results presented here are evaluated in this scenario. The unscaled parameters for the baseline case are as follows: $t_0 = 0, t_{max}=100 R_0/v_{th}, N_p=10^4, N_c=2\times 10^2, N_{real}=10, N_t=2\times 10^3, T_i=0.8keV, T_e=1.5keV, B_0=3.7T, R_0=2.5 m, L_{T_i}=R_0/3, L_{T_e}=R_0/3, Z_{eff}=2, A_{eff}=3, n_0=5\times 10^{19} m^{-3}, a_0 = 0.5m, \Omega_t = 0, c_1=1, c_2=0, c_3=3, \Phi = 1\% , A_i = 0.9, \lambda_x = 5\rho_i, \lambda_y=5\rho_i, \lambda_z = R_0, \tau_c = R_0/v_{th}, k_{0i} = 0.1\rho_i^{-1}, k_{0e}=0.1\rho_i^{-1}, r_0=0.35m, T=T_i, A=1, Z=1$. Unless otherwise stated, the initial distribution function of the test particles is assumed to be a local Maxwellian with temperature $T$, localized across a flux tube at $x = r_0$.

\subsection{Single particle testing}
\label{Section_4.1}

We begin by evaluating whether the code can capture basic invariants and analytical results at the level of individual particle trajectories. We first analyze purely neoclassical motion, followed by turbulent scenarios.

\begin{figure}
	\subfloat[\label{fig_1}]{
	\includegraphics[width=0.45\linewidth]{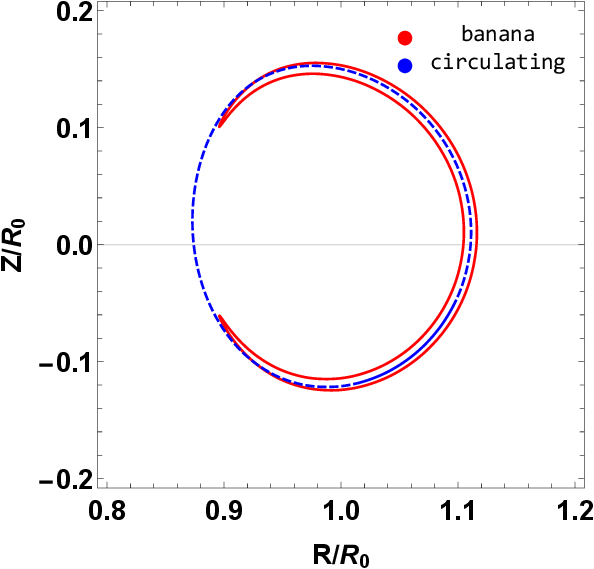}
	}	
	\subfloat[\label{fig_1_t}]{
	\includegraphics[width=0.52\linewidth]{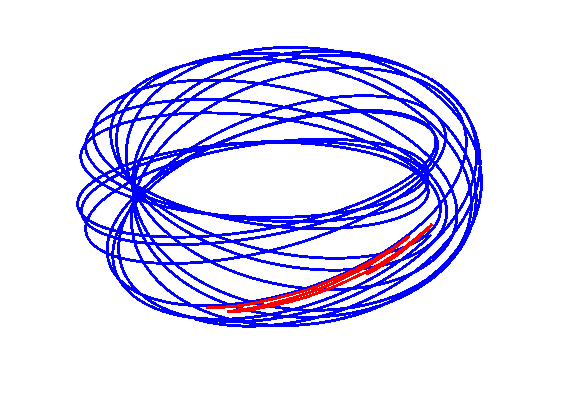}
	}
	\caption{Typical banana (red, full line) and circulating (blue, dashed) trajectories obtained for purely neoclassical motion in the baseline scenario in poloidal projection a) and full 3D representation b). }
\end{figure}

The fundamental requirement is that the two types of unperturbed neoclassical trajectories—passing (circulating) and trapped (banana)—are accurately reproduced. This is confirmed in Fig. \eqref{fig_1}, where the poloidal plane projection $(R,Z)$ of two typical neoclassical trajectories in realistic equilibria of WEST $\# 54178$ is shown. The curves are perfectly closed, indicating that the dynamics is accurately captured over multiple bounce or passing times. Additionally, these trajectories resemble the flux surfaces that are not perfectly circular but exhibit triangularity and elongation. Fig. \eqref{fig_1_t} displays the same trajectories in full 3D geometry.

\begin{figure}[h!]
	\centering
	\includegraphics[width=0.5\linewidth]{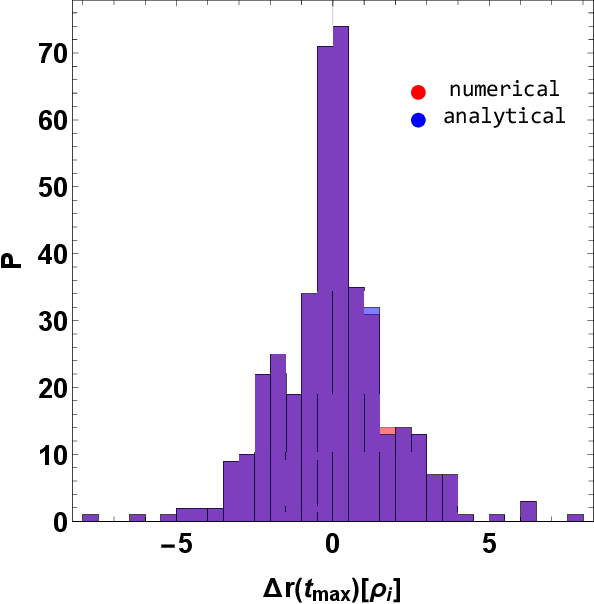}
	\caption{Distribution of $\Delta r(t_{max})$ for an ensemble of particles obtained with the T3ST code (red) against corresponding distribution obtained with the analytical formula \eqref{eq_4.1.1}.}
	\label{fig_2}
\end{figure}

While the emergence of trapped and passing trajectories provides a qualitative validation of our integrator, quantitative assessments are also essential. For this purpose, we analyze the baseline scenario using a circular equilibrium model 
\eqref{Section_2.2.1}, excluding collisions and turbulence to ensure the preservation of motion invariants such as the Hamiltonian (energy) $E=q\Phi^\star$, toroidal momentum $P_c$ and magnetic moment $\mu$. 

\begin{figure}
	\subfloat[\label{fig_3}]{
		\includegraphics[width=.49\linewidth]{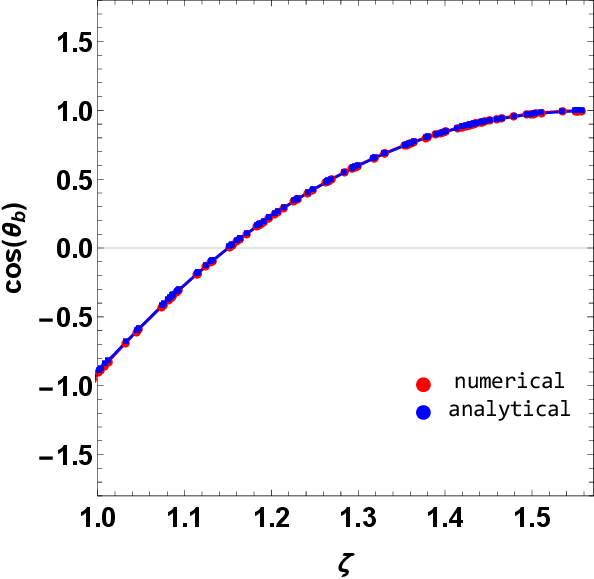}%
	}	
	\subfloat[\label{fig_4}]{
		\includegraphics[width=.49\linewidth]{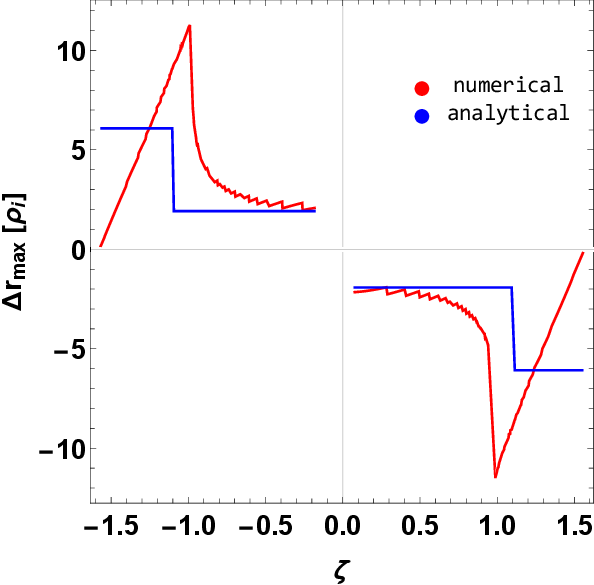}%
	}
	\caption{Bounce angles (a) and orbit widths (b) obtained with T3ST (red) versus analytical estimations \eqref{eq_4.2} (blue) for the baseline scenario in the limit of circular equilibrium. }
\end{figure}

Particles are initialized at $r_0$ and allowed to move under neoclassical constraints. We evaluate their radial displacement $\Delta r(t)$. For a circular magnetic equilibrium, a first-order analytical estimation of $\Delta r(t)$ can be derived, using the linearization $\psi(r)\approx \psi(r_0)+\Delta r \psi'(r_0)$, as a function of invariants and  parallel velocity $v_\parallel(t)$:
\begin{align}\label{eq_4.1.1}
	\Delta r (t) \approx \frac{\mu m }{q b_\theta(r_0)}\left(\frac{v_\parallel(t) }{E-\frac{m v_\parallel^2(t)}{2}}-\frac{v_{\parallel}(0) }{E-\frac{m v_{\parallel}^{2}(0)}{2}}\right) .
\end{align}
We compare this analytical result againts numerical data for $200$ test-particles in Fig. \eqref{fig_2} and find almost perfect agreement. 

\begin{figure}
	\subfloat[\label{fig_5}]{
		\includegraphics[width=.57\linewidth]{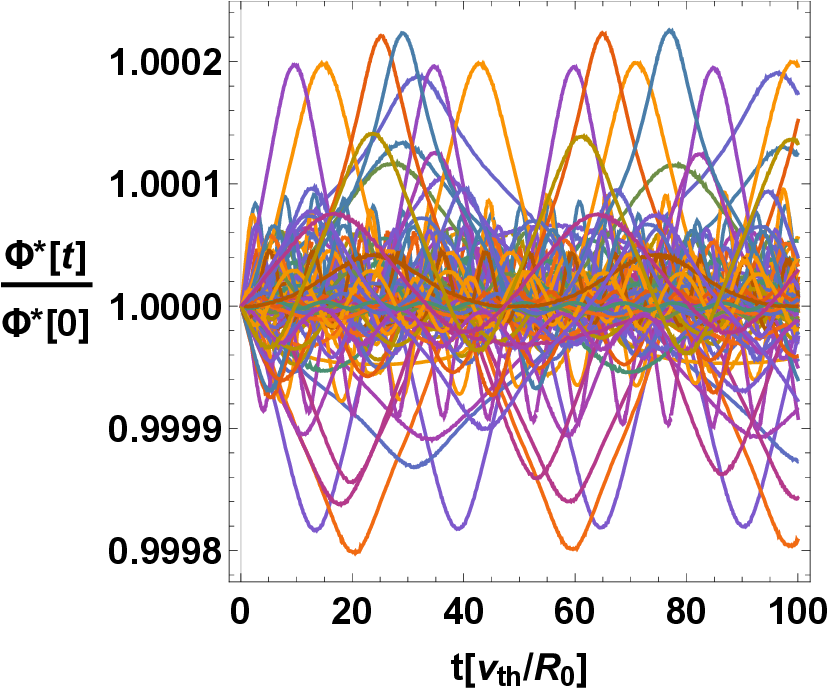}%
	}	
	\subfloat[\label{fig_6}]{
		\includegraphics[width=.44\linewidth]{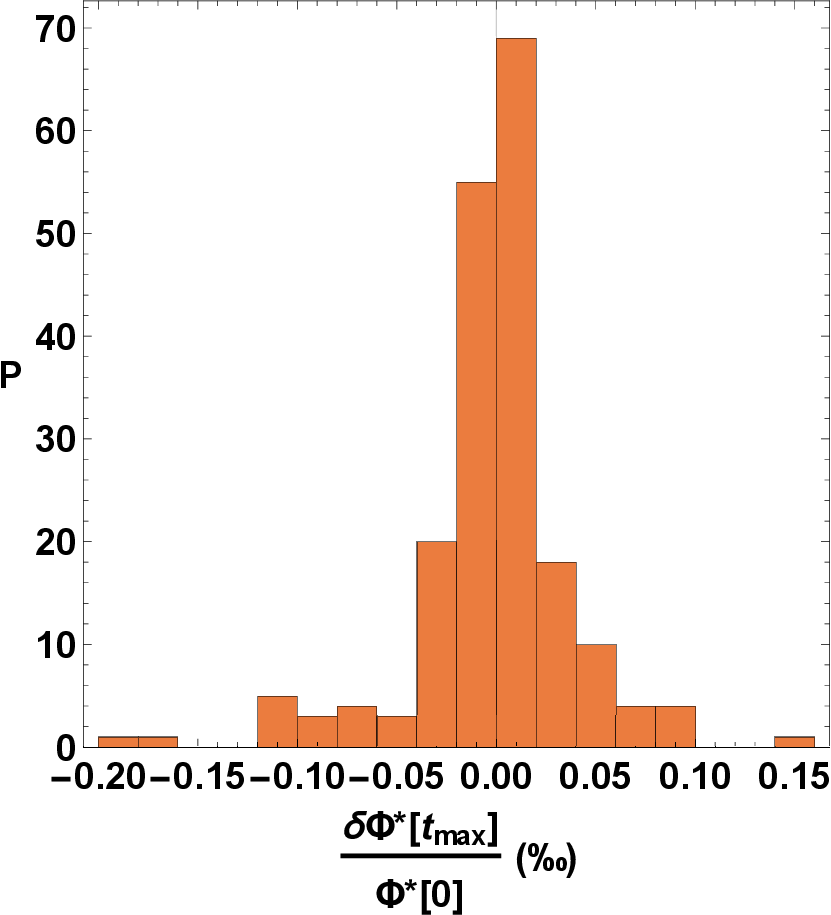}%
	}
	\caption{Time evolution of the relative Hamiltonian function (a) and distribution of relative errors at the end of the computing time $t_{max}$.}
\end{figure}

The bounce angle $\theta_b$ (maximum poloidal angle of a banana trajectory) and the orbit width $\Delta r_{max}$ can also be estimated analytically from initial conditions ($\tan \zeta = v_\perp/v_\parallel$):
\begin{align}\label{eq_4.2}
	\cos\theta_b &\approx \frac{R_0}{r_0}\left(\frac{\mu B_0}{E}-1\right)\\
	\Delta r_{max} &\approx \rho_i\bar{q}(r_0)\frac{B(r_0)}{B}\sqrt{\frac{A}{Z}}\left(1+\frac{r_0}{R_0}\right), |\zeta|\to 0\\
	\Delta r_{max} &\approx \rho_i\bar{q}(r_0)\frac{B(r_0)}{B}\sqrt{\frac{A}{Z}}\left(1+\frac{r_0}{R_0}\right)\sqrt{\frac{R_0}{r_0}}, |\zeta|\to\pi/2.
\end{align}

Comparisons between numerical results and analytical estimations for these quantities are shown in Figs. \eqref{fig_3},\eqref{fig_4}. In contrast with the bounce angles, the orbit widths show only an approximate agreement with analytical values since the latter are derived for asymptotic cases.
\begin{figure}[h!]
	\centering
	\includegraphics[width=0.75\linewidth]{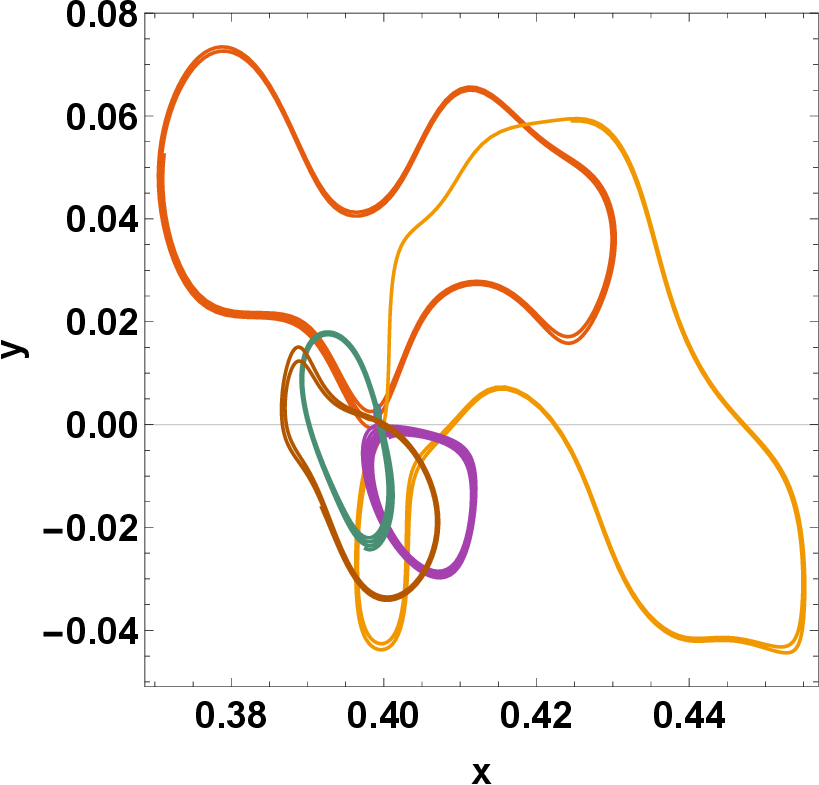}
	\caption{Stochastic trajectories of cold $E_{kin}=0$ particles in the baseline scenario and frozen turbulence.}
	\label{fig_7}
\end{figure}

We further investigate the conservation of energy for particle ensembles. While Figs. \eqref{fig_1}-\eqref{fig_4} indirectly confirm conservation, we explicitly test it for a set of $200$ test particles. Fig. \eqref{fig_5} illustrates the relative variation of the Hamiltonian $\Phi^\star$ over time, showing oscillations on the order of $10^{-2}\%$ without any secular growth. Fig. \eqref{fig_6} shows a histogram of numerical errors at $t_{max}$, demonstrating stability even in a realistic GEQDSK equilibrium that require interpolation.

\begin{figure}
	\subfloat[\label{fig_8}]{
		\includegraphics[width=.56\linewidth]{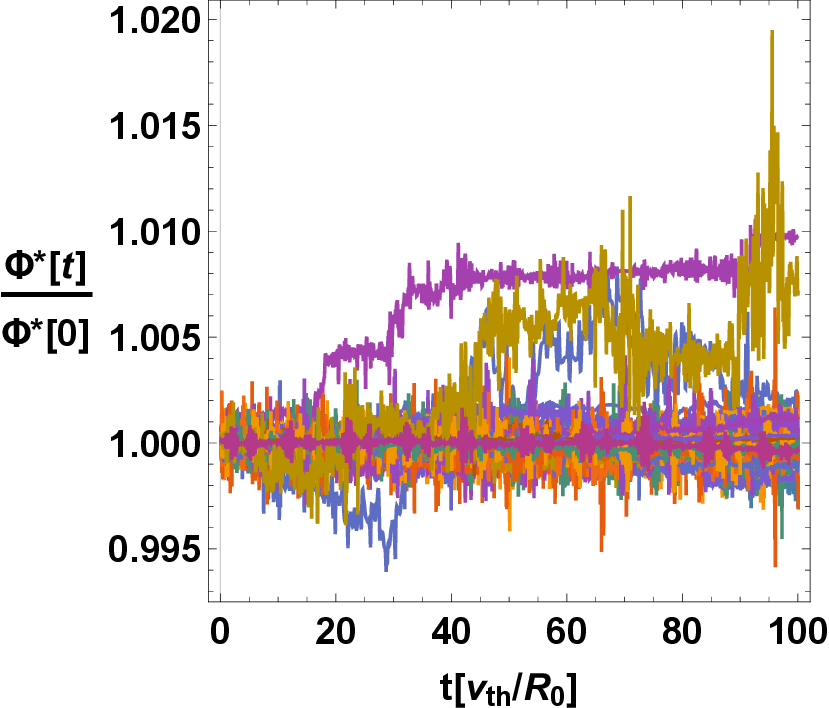}%
	}	
	\subfloat[\label{fig_9}]{
		\includegraphics[width=.44\linewidth]{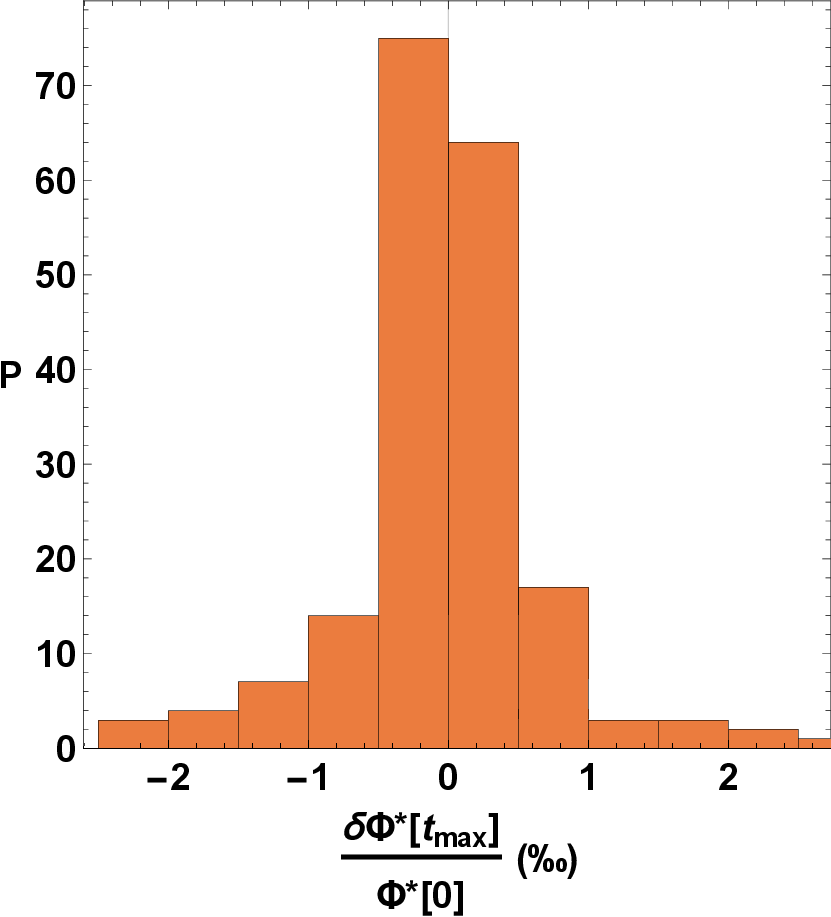}%
	}
	\caption{Time evolution of the relative Hamiltonian function (a) and distribution of relative errors at the end of the computing time $t_{max}$ for purely turbulent trajectories.}
\end{figure}

For the ITG turbulence case, we begin with a qualitative test by setting the turbulent electrostatic field $\phi_1(x,t)$ to "frozen" ($\omega=0$ for all modes). Additionally, only "cold" particles ($E=0$) are used. This ensures purely turbulent dynamics, with negligible magnetic drifts and only a small parallel acceleration, resulting in approximately closed 2D-Hamiltonian trajectories. Fig. \eqref{fig_7}
illustrates this behaviour in the field-aligned space $(x,y)$. Energy conservation is tested under frozen turbulence, as shown in Figs. \eqref{fig_8} and \eqref{fig_9}. Although numerical errors grow by nearly two orders of magnitude in comparison with pure neoclassical motion, they remain within $\sim 0.1\%$, validating the integrator’s robustness under these challenging conditions.

\subsection{Neoclassical transport testing}
\label{Section_4.2}

The code's ability to simulate particle distributions is critical for transport studies. To validate this, we consider the baseline scenario with a circular magnetic equilibrium. Maxwell-Boltzmann distributed particles with uniform pitch angles are initialized at a single spatial point $r=r_0,\theta = \pi/2$, which corresponds to $R=R_0,Z=r_0$ or $x=r_0, y=r_0 \pi/2, z=\pi/2$ (above magnetic axis). Particle transport coefficients \eqref{eq_2.7.2},\eqref{eq_2.7.3}, at small times are defined as $V_r^{t=0} \equiv V_n(0|r_0),D_r^{t=0} \equiv D_n(0|r_0)$.

\begin{figure}
	\subfloat[\label{fig_10}]{
		\includegraphics[width=.51\linewidth]{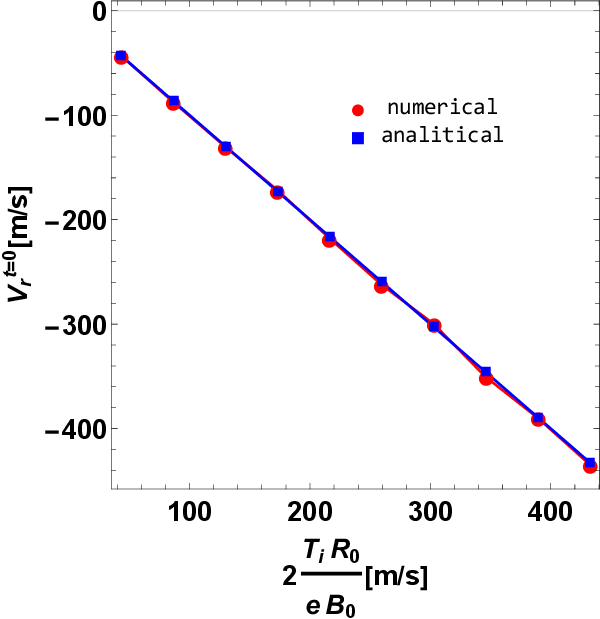}%
	}	
	\subfloat[\label{fig_11}]{
		\includegraphics[width=.49\linewidth]{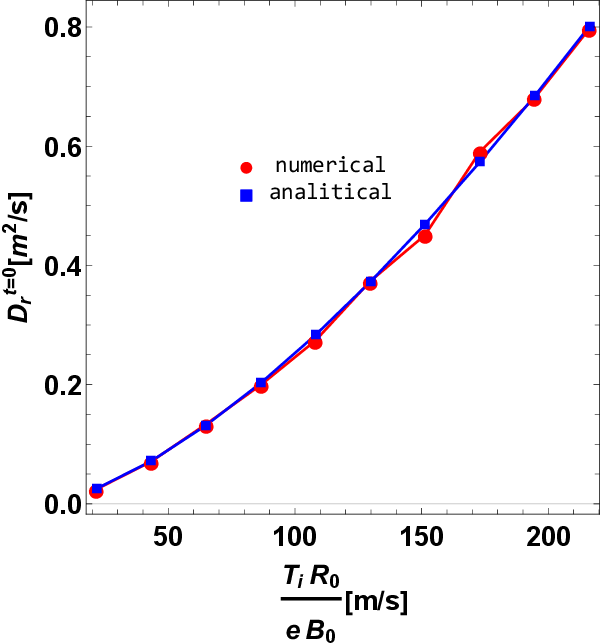}%
	}
	\caption{Initial radial velocity (a) and diffusion (b) computed with T3ST (red, circle) and analytical formulae \eqref{eq_4.2.2} (blue,dots) for a Maxwellian distribution of test-particles placed at $r=r_0,\theta=\pi/2$ in the baseline scenario at different temperatures.}
\end{figure}

This is an interesting case since we can evaluate analytically $V_r^{t=0},D_r^{t=0}$. We start from the eqns. \eqref{eq_2.1.2a} and projects the time derivatives on the radial direction. By using an expansion in the small gyrokinetic parameter $\rho_i\nabla$, the low-beta approximation $\nabla\times\mathbf{B} \approx 0$, the circular model and the condition $\theta=\pi/2$, we get:
\begin{align}\label{eq_4.2.1}
\dot{r} \approx -\frac{m(v_\parallel^2+v_\perp^2/2)}{qB^3}\left(\nabla B\times\mathbf{B}\right)\cdot e_r\approx -\frac{m(v_\parallel^2+\mu B)}{qB R}.
\end{align}

After thermal averaging over the MB distribution, we obtain:
\begin{align}\label{eq_4.2.2}
	\langle\dot{r}\rangle &\approx -\frac{2T}{qB_0 R_0} = V_r^{t=0}\\
    \langle\dot{r}^2\rangle - &\langle\dot{r}\rangle^2 \approx 3\frac{R_0}{v_{th}}\left(\frac{T}{q B_0 R_0}\right)^2  = \frac{D_r^{t=0}}{t}.
\end{align}

We plot the numerical results againts these analytical values in Fig. \eqref{fig_10},\eqref{fig_11} and find very good agreement, thus, validating both the numerical evaluation of drifts, as well as the thermal distribution of particles. 

\begin{figure}[h!]
	\centering
	\includegraphics[width=0.75\linewidth]{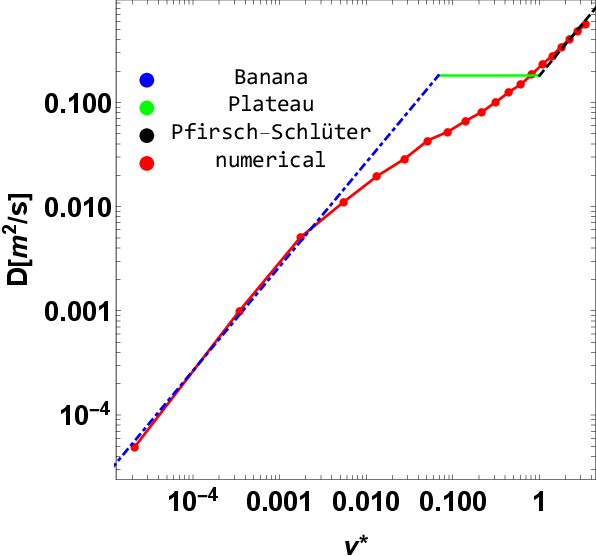}
	\caption{Numerical (red) against theoretical (blue,green,black) diffusion coefficient for particle neoclassical transport in the baseline scenario for $H$ ions and circular equilibrium.}
	\label{fig_12}
\end{figure}

A true verification of T3ST's ability to predict transport is the neoclassical case. We begin by introducing several key parameters: the inverse aspect ratio $\varepsilon=r_0/R_0$, the thermal velocity $v=\sqrt{2T/m}$, particle's Larmor radius $\rho = m v/qB$, the orbit transit frequency $\Omega_{orb}=v\sqrt{\varepsilon}/(\bar{q}(r_0)r_0\sqrt{2})$, the real $\nu$ and normalized $\nu^\star=\nu/\Omega_{orb}$ collisional frequencies and $D_C=\rho^2\nu/2$, the classical diffusion. There are well-established regimes of neoclassical transport (\cite{Hirshman_1981}): the low-collisional "banana regime", valid at $\nu^\star\ll \varepsilon^{3/2}$ where the particle diffusion coefficient is $D_B=\varepsilon^{-3/2}\bar{q}(r_0)D_C$; the high-collisional Pfirsch-Schluter regime $\nu^\star\gg 1$ where $D_{PS}=\bar{q}^2 D_C$;  the intermediate "plateau regime", $\nu^\star\sim 1$, with $D_P=\bar{q}^2 D_C\nu^\star$. 

To validate the code's ability to simulate neoclassical transport, the baseline scenario is extended by introducing collisions via the Lorentz operator \eqref{eq_2.3.1}. The study focuses on $H$ ions in a circular equilibrium, as the analytical neoclassical diffusion coefficients are derived for this configuration. The collisional frequency $\nu$ is varied systematically, and the resulting numerical diffusion coefficients are compared with analytical predictions. The results, depicted in Fig. \eqref{fig_12}, demonstrate strong agreement between the numerical simulations and theoretical values, confirming the code's capability to accurately model neoclassical transport across different collisionality regimes.

\subsection{Turbulent transport testing}
\label{Section_4.3}

\begin{figure}
	\subfloat[\label{fig_18}]{
		\includegraphics[width=.51\linewidth]{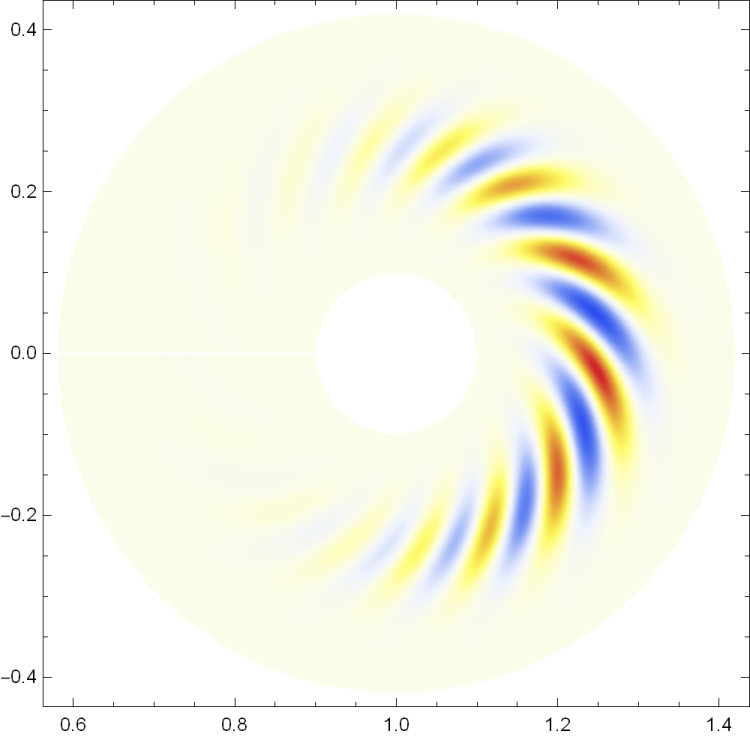}%
	}	
	\subfloat[\label{fig_19}]{
		\includegraphics[width=.5\linewidth]{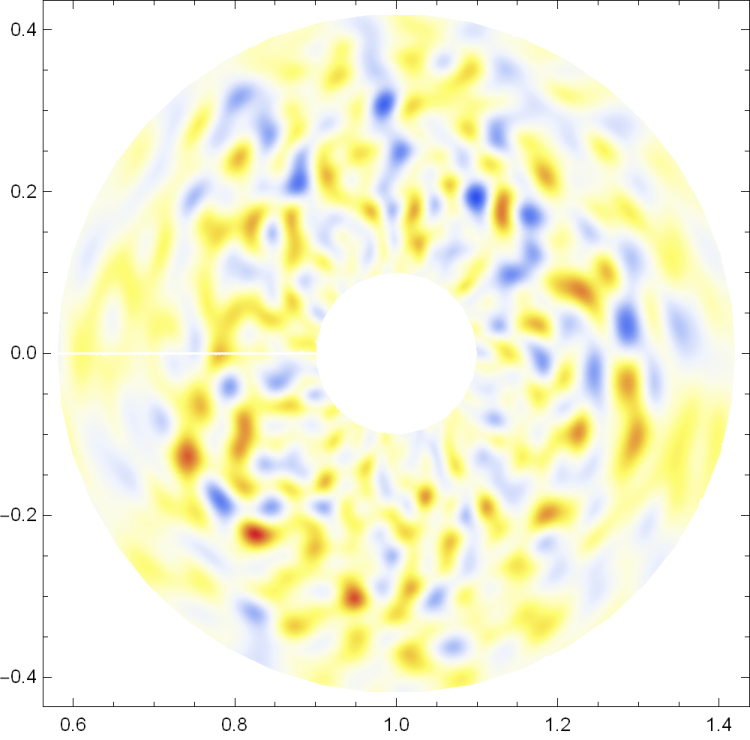}%
	}
	\caption{a) The most probabale Fourier mode of an ITG turbulent field, $\phi_1^{\bar{\mathbf{k}}}(r,\theta)$, where $\bar{\mathbf{k}}\equiv \left(\langle k_x^2\rangle^{1/2},\langle k_y^2\rangle^{1/2},\langle k_z^2\rangle^{1/2}\right)$ modulated by a localized radial and parallel profile $exp(-(r-r_0)^2-z^2)$. b) Full electrostatic potential $\phi_1(r,\theta)$ in fully developed turbulence.}	\end{figure}

We move further to the problem of turbulent transport and start by checking the quality of synthetic turbulence. We do that by examining a single ITG Fourier mode and fully developed ITG states as being generated within T3ST, in the simple case of circular geometry and poloidal projection. One can define the most probable mode as being described by the average wavevector $\bar{\mathbf{k}}\equiv \left(\langle k_x^2\rangle^{1/2},\langle k_y^2\rangle^{1/2},\langle k_z^2\rangle^{1/2}\right)\approx \left(0.2\rho_i^{-1},0.4\rho_i^{-1},0.5\right)$, provided that the spectrum \eqref{eq_2.4.2} has been used and the parameters took the baseline case values. In Fig. \eqref{fig_18} it can be seen the poloidal frame of this mode, modulated by a balloning radial and parallel envelope $exp(-(r-r_0)^2-z^2)$. Obviously, it resembles typical eigenmodes of the gyrokinetic system of equations in the linear limit. Fully developed turbulence as a superposition of many $(N_c)$ individual modes that lose the balloning property is shown in Fig. \eqref{fig_19}.

We move now to the question of weather the ensemble representation of turbulent fields is accurate at the level of trajectories. For that, the "most turbulent case" is chosen, where all particles are cold $T=0$ and localized at the low-field-side equatiorial point $r=r_0,\theta=0$. Their motion is computed for a short time interval in a large number of field realizations ($N_{real} = 10^4$). We note that, similarly with the discussion regarding pure neoclassical motion, we can estimate analytically the diffusion in this simplified case.  

\begin{figure}
	\centering
	\includegraphics[width=0.75\linewidth]{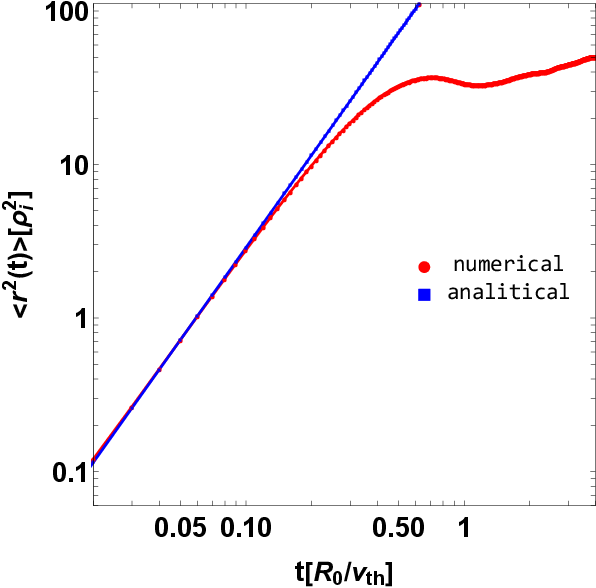}
	\caption{Small time spreading of cold $H$ ions located at the equatorial low-field-side of the baseline scenario. Comparison between numerical results (red, bullet) and analytical estimation (blue, square).}
	\label{fig_13}
\end{figure}

Since our particles are cold at the low-field-side, initially there are no magnetic drifts and the sole component of motion is, in fact, the ExB drift, i.e. $\dot{\mathbf{X}} = \mathbf{v}=-\nabla \phi_1^{gc}\times \mathbf{b}/B$. Since we are interested in radial transport, i.e. along the radial coordinate $x$, we have $\dot{x}(t) = \mathbf{v}\cdot\nabla x = v_x(t)\approx -\partial_y \phi_1/B \nabla x\cdot\left(\nabla y\times\mathbf{b}\right)$. In circular magnetic geometry, $v_x(t=0)\approx -\partial_y \phi_1(\mathbf{0},0)/(\bar{q}B_0)$. The statistical average of this velocity dictates the initial radial particle spreading $\langle r^2(t)\rangle -r_0^2 = t^2 \langle v_x^2(0)\rangle$. It is straightforward to show, based on the Fourier representation of fields, that $\langle v_x^2(0)\rangle = \langle \phi_1^2(0)\rangle \langle k_y^2\rangle/(\bar{q}B_0)^2=\left(\Phi v_{th}\rho_i \sqrt{3}/\lambda_y\right)^2$. In fig. \eqref{fig_13} we compare this analytical estimation with the numerical results obtained with T3ST and find very good agreement at $t\ll 1$. 
 
\begin{figure}
	\centering
	\includegraphics[width=0.75\linewidth]{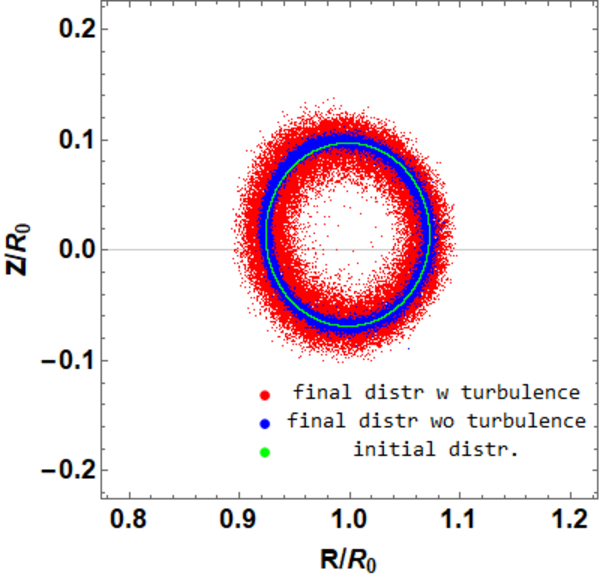}
	\caption{Initial (green), asymptotically neoclassical (blue) and turbulent at $t_{max}$ (red) positions of test-particle projected onto the poloidal plane for the baseline scenario (WEST discharge $\#54178$).}
	\label{fig_14}
\end{figure}

Although not explicitly a validation of our code, we believe there is a need to visualize the transport of test particles as predicted by T3ST. In Fig. \eqref{fig_14}, we show the positions of many gyrocenters, color-coded in green, blue, and red, projected onto the poloidal plane $(R, Z)$. The green dots show an almost elliptical distribution, which corresponds to the initialization of test particles across a flux surface. Together with the Maxwell-Boltzmann assumption, this defines the entire kinetic distribution function $f_0$. Unfortunately, this is not a canonical Maxwellian, thus, not a true equilibrium solution of the GK equation. Consequently, when the particles are allowed to move without collisions or turbulence, they follow their neoclassical trajectories, leading to a dense, relatively thick annulus, which can be seen in blue in Fig. \eqref{fig_14}. The blue state is reached asymptotically, representing an equilibrium state where all radial transport is zero, but with a non-zero width related to finite Larmor radius effects. Finally, the inclusion of turbulence (or collisions, for that matter) induces radial transport. This means that the particles no longer remain on fixed, confined trajectories but instead experience stochastic jumps between various turbulent potential lines, resulting in consistent transport. This is represented by the red distribution in Fig. \eqref{fig_14}, showing the position of particles in the presence of turbulence at the end of the simulation time $t_{max}$.

\begin{figure}
	\subfloat[\label{fig_15}]{
		\includegraphics[width=.51\linewidth]{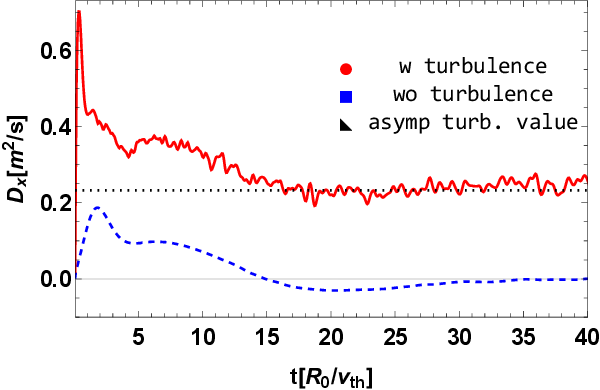}%
	}	
	\subfloat[\label{fig_16}]{
		\includegraphics[width=.5\linewidth]{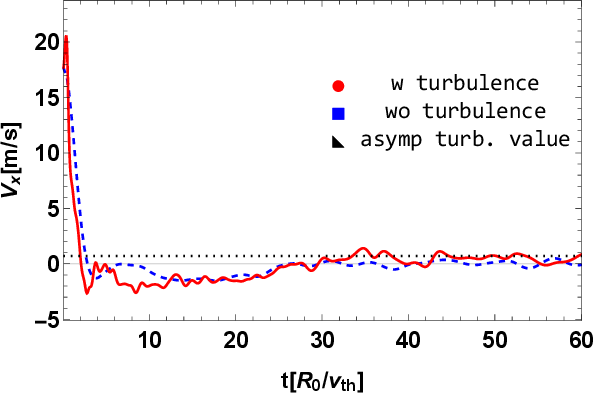}%
	}
	\caption{Radial transport in the baseline scenario with (red) and without (blue) turbulence. Dotted line shows asymptotic turbulent value.}
\end{figure}

We ilustrate this behavior by plotting in Fig. \eqref{fig_15},\eqref{fig_16} the running particle transport coefficients $V_n(t|x=x_0),D_n(t|x=x_0)$ for the purely neoclassical case (blue, dashed) and for the turbulent case (red, full line). We also add, in dotted-black, the asymptotic values (experimentally relevant) the turbulent case, $V_n(x_0),D_n(x_0)$. One must note that the percolation of the initial Maxwellian distribution to confined neoclassical trajectories leads to zero transport coefficients at long times. In contrast, with turbulence, we get finite valued transport.

\subsection{Validation and illustration}
\label{Section_4.4}

In this section we indend to validate T3ST against gyrokinetic simulations of an intricate problem in order to emphasize both its versatility and inaccuracies. We choose the problem of energetic ion turbulent transport driven by drift-type microinstabilities and compare our data with results from literature \cite{PhysRevLett.101.095001}.

The physical scenario is the so-called "Cyclone base case" corresponding to a typical DIII-D discharge \cite{A.M.Dimits_2000}. This is standard for benchmarking any newly developed gyrokinetic code and has been extensively investigated in the past two decades \cite{Falchetto_2008}. The physical parameters relevant for T3ST are: $T_i=T_e=0.5keV, B_0 = 1.9T, R_0=1.71m, R_0/L_{T_i} =6.9, R_0/L_n=2.2, a= 0.625m, c_1 = 0.85, c_2 = 0, c_3 = 2.2, r_0 = a/2$. For H ions, $\rho_i/a=\rho^\star \approx 1/519$. The magnetic equilibrium is considered circular and the parameters $c_1,c_2,c_3$ are chosen such that $\bar{q}(r) = c_1+c_2	(r/a)^1+c_3	(r/a)^2$, thus, $\bar{q}(r_0) = 1.4, \hat{s} = 0.78$. The plasma is subject to the ITG instability which has been show \cite{10.1063/1.1647136}, via GK simulations, to have linear modes with an approximately constant phase velocity $v_{ph}\approx V_\star \approx v_{th}\rho_i/L_n $, i.e. $\omega_{\mathbf{k}} \approx v_{ph}k_\theta$, while the growth rates $\gamma$ encompass the $[0,0.7]\rho_i^{-1}$ interval with a maxima at $k_\theta\rho_i\approx 0.3$. The ITG instability saturates at large times into a turbulent state with a relative intensity at midradius $r_0$ of $\Phi = eA_\phi/T_i\approx 1.1\%$, a radial correlation length of $\lambda_r \approx 7\rho_i$ and a peaked spectrum in the poloidal wavenumber at $k_\theta\rho_i\approx 0.15$.  
Moreover, the time correlation of the electrostatic potential is\cite{10.1063/1.1647136} $\langle\phi_1(\mathbf{x},t)\phi_1(\mathbf{x},0)\rangle \propto \exp\left(-t/\tau_c\right)$ where the correlation time was found to be $\tau_c\approx 1/\gamma\approx 3/\omega\approx 10\rho_i/v_{ph}$. For T3ST, these parameters translate into scaled values as $A_i = 1, \Phi =0.011,\lambda_x=7,\lambda_y=5,k_0=0.05,\tau_c = 4$ while we choose $\lambda_z=1$.

The main quantity of interest is the phase-space (kinetic energy-pitch angle, $E-\lambda$) structure of the particle diffusion coefficient defined here as $D(E,\lambda) = \lim\limits_{t\to t_{max}} \langle (r(t)-r(0))^2\rangle/2t$, $t_{max}=60R_0/v_{th}$. We note that this is only an approximation for the true diffusion coefficient \eqref{eq_2.7.2}, but we employ it for consistency with literature \cite{PhysRevLett.101.095001}. For the same reason, the particles are initially distributed, not on the $r_0$ flux surface, but on a thick anulus $r\in (0.45,0.55)a$ with fixed kinetic energy $E$ and pitch angle $\lambda$. 

\begin{figure}
	\includegraphics[width=.8\linewidth]{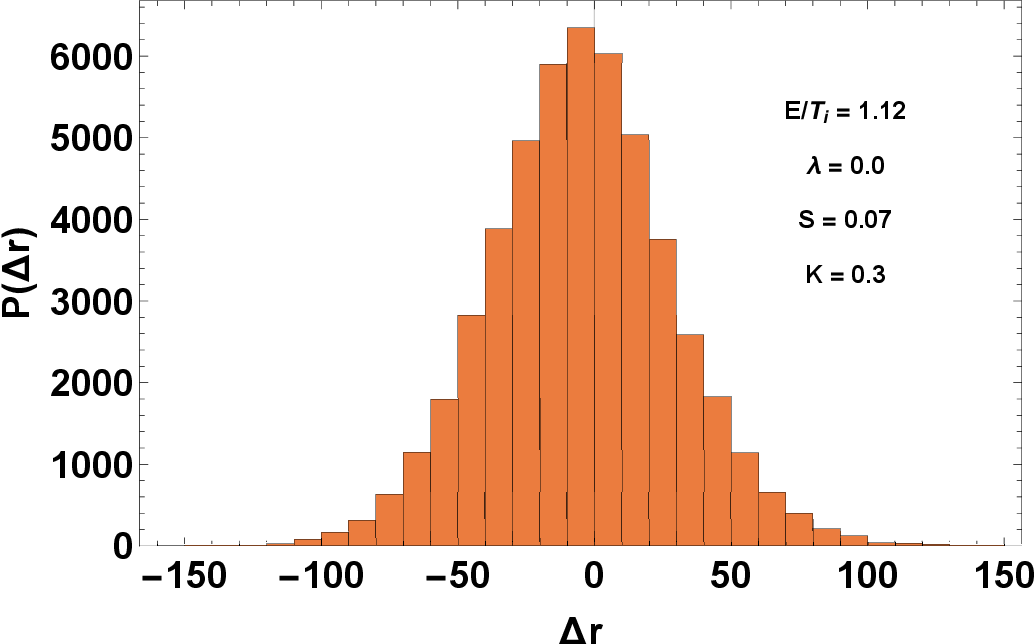}%
	\caption{PDF of radial displacements $\Delta r/\rho_i$ of particles at $t_{max}=60 R_0/v_{th}$. All particles have $E=1.7T_i$, an initial $\lambda=0$ pitch angle. The PDF has skewness $S$ and kurtosis $K$.}
	\label{fig_18_a}
\end{figure}

The first check is that, in accordance with the results from \cite{PhysRevLett.101.095001}, the $\rho^\star \approx 1/500$ limit makes the transport proces close to diffusive. That can be seen from the distribution of radial displacements of particles in Fig. \eqref{fig_18_a} which remains close to normal at the final computing time $t_{max}$. 

Second, we look at the total particle diffusion coefficient that can be obtained from the integration over the phase-space $D_{eff} = \int dE d\lambda P(E,\lambda) D(E,\lambda)$, where the kinetic distribution is Maxwell-Boltzmann $P(E,\lambda) = \sqrt{E/T_i}\exp\left(-E/T_i\right)$. T3ST provides a value for this quantity of $D_{eff}\approx 2.9 \chi_{GB} $ that can be compared with results from \cite{PhysRevLett.101.095001}, $D_{eff}\approx 2.2\chi_{GB}$. It seems that T3ST overestimates the diffusion coefficient by $30\%$ in this case.

\begin{figure}
	\includegraphics[width=.8\linewidth]{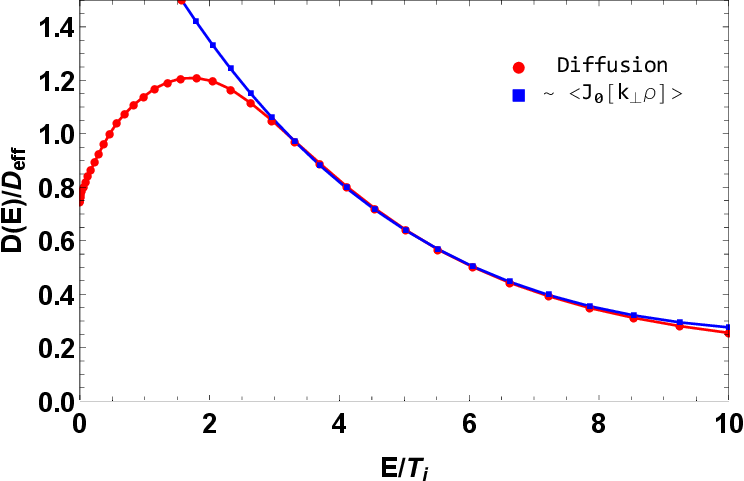}%
	\caption{Energetic dependence of diffusion.}
	\label{fig_18_b}
\end{figure}

We move to the dependence of radial diffusion on particle energy, that means integrated over the distribution of pitch angles $D(E) = \int d\lambda D(E,\lambda)$. We plot the results obtained by T3ST in Fig. \eqref{fig_18_b}. We observe an increase in transport at small energies, a peak and a slow, algebraic, decay after that. These results are similar with the ones obtained in \cite{PhysRevLett.101.095001} with the main difference that our peak is not located at $E/T_i = 2$ but at $E/T_i\approx 1.7$. Otherwise, the peak value relative to the effective diffusion is $\approx 1.2$ comparable to the $\approx 1.4$ value reported by GKs. The reason for the decay at larger energies is well known: energetic particles do not "see" the structure of turbulence directly, but in an averaged manner, first as gyro-average \cite{hauff0} and second, arguably, as bounce-time avearge \cite{10.1063/1.3013453,PhysRevLett.101.095001}. This effectively makes the particles to feel an effective turbulent amplitude that goes like $\Phi^\prime\sim \Phi J_0(k_\perp\rho(E))$ where $J_0$ is the Bessel function and $\rho(E) = m v_\perp/qB\sim \sqrt{E}$. This behavior is indeed exhibited by our results as seen in Fig. \eqref{fig_18_b}. 

\begin{figure}
	\includegraphics[width=.8\linewidth]{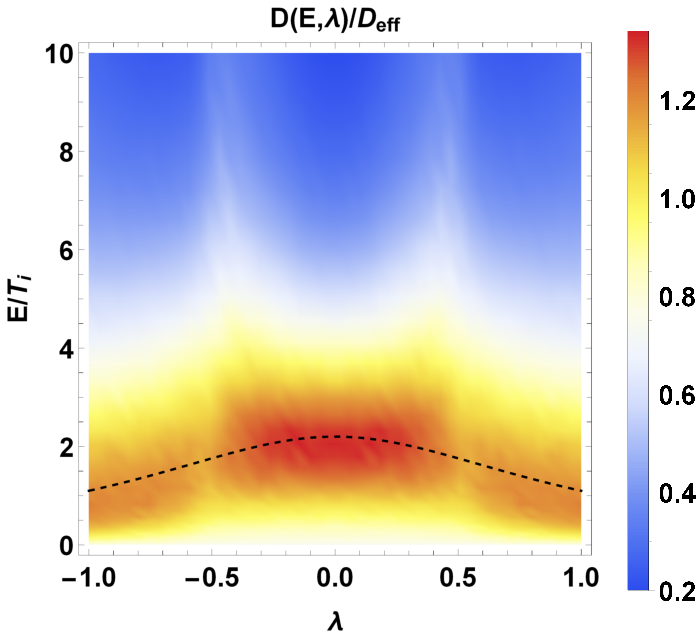}%
	\caption{Phase-space structure of the diffusion.}
	\label{fig_18_c}
\end{figure}

The existance of a transport resonance in the $E-\lambda$ phase space can be understood as a particle-wave resonance \cite{Dong_2022}. The ITG turbulent potential moves, approximately, along the "bi-normal" direction $\nabla y$ with the phase velocity $v_{ph}\approx \rho_i v_{th}/L_n$. Particles, on the other hand, have magnetic drifts that can be evaluated on the $y$ direction as $v_d \approx -E/T_i (1+\lambda^2)\rho_i v_{th}/R_0$. 
If $v_d\approx v_{ph}$ then particles resonate with the drift-wave and remain correlated for a longer time, experiencing transport. In contrast, when the particle can't keep up with the ITG drift, it decorrelates and the transport decays. Therefore, it is natural that maxima in diffusion are to be found at $v_d\approx v_{ph}$ which implies $E(1+\lambda^2)/T_i\approx R_0/L_n\approx 2.2$. 	

This behavior is apparent when visualising the phase-space structure of the diffusion coefficient $D(E,\lambda)$ shown in Fig. \eqref{fig_18_c}. There, the black-dotted line indicates the $2.2/(1+\lambda^2)$ curve that seems to pass across the local maxima of $D(E,\lambda)$ which is in accordance with the mechanism of particle-wave resonance. The same behavior is apparent in the GK results \cite{PhysRevLett.101.095001}. Supplementary, there is an obvious departing behaviour at $\lambda\approx 0.5$ between the transport of trapped ($\lambda<0.5$)  and passing ($\lambda>0.5$) particles. However, T3ST predicts high levels of diffusion in the $|\lambda|\to 1$ limit, in contrast with the GK data. 

Most likely, the quantitative differences between the results of T3ST and gyrokinetics \cite{PhysRevLett.101.095001} are due to an insufficiently detailed representation of the real ITG turbulence, in particular across the parallel coordinate $z$. Nonetheless, through all similitudes, T3ST has proved itself as a reliable numerical tool that can characterize the turbulent transport both qualitatively and quantitatively in good agreement with more sophisticated methods such as gyrokinetics. 

\section{Conclusions and perspectives}
\label{Section_5}

In the present work, we have introduced the Turbulent Transport in Tokamaks via Stochastic Trajectories (T3ST) code. Its development is motivated by the existence of a gap in the current landscape of numerical codes used by the nuclear fusion community for studying turbulent transport. This gap arises from the need for a tool that is non-linear, accurate, versatile with respect to the properties of turbulence, and computationally efficient. These requirements are not fully met by gyrokinetic codes, which are extremely computationally intensive, nor by simplified quasilinear approaches, which lack versatility and the non-linear dynamical character.

T3ST fulfills these requirements by employing the test-particle approach, combined with a statistical description of turbulence in terms of ensembles of synthetic random fields with prescribed properties. The neoclassical dynamical components are accurately reproduced, as the code can handle realistic magnetic equilibria while incorporating various relevant effects, such as magnetic drifts, toroidal rotation, and centrifugal effects.

The main drawback of T3ST is its reliance on input parameters, such as correlation lengths and turbulence strengths, due to the omission of self-consistent turbulence evaluations. This limitation makes it less comprehensive than gyrokinetic codes. However, this limitation is also its strength, as the avoidance of plasma-field equation computations results in significantly lower computational requirements. As a result, T3ST is orders of magnitude faster than gyrokinetic simulations. Additionally, the statistical ensemble representation of turbulence provides the code with straightforward means of exploring the dependence of transport on various physical regimes. This level of flexibility is challenging to achieve in self-consistent approaches, where the resulting fluctuations are heavily correlated with numerous plasma parameters.

In Section \eqref{Section_2}, we provided a detailed description of all the equations of the model, ranging from the equations of motion to collisional operators, possible magnetic equilibria, and other key aspects. Section \eqref{Section_3} discussed the numerical implementation, including details on coordinate systems, scaling, particle sampling, trajectory integration, and the stochastic generation of fields. Finally, in Section \eqref{Section_4}, we presented numerical tests to evaluate the accuracy of the code and validate it through comparisons with results from the literature. These tests demonstrated that T3ST is both fast and accurate across all analytical cases. Furthermore, the code successfully describes the turbulent transport, yielding results consistent with existing data.

With the development of a functional and tested version of the code, the authors foresee three primary avenues for future work. First, a comprehensive validation campaign against gyrokinetic simulations and experimental data is a crucial next step in the evolution of T3ST. Second, further refinements and extensions to the code are both possible and, potentially, necessary. For example, one could incorporate more sophisticated and realistic turbulence spectra, include electron-temperature-gradient (ETG) effects for electron studies, account for magnetic fluctuations (including resonant magnetic perturbations, or RMPs), or implement additional particle scenarios, such as beam ions. Finally, T3ST is well-positioned to investigate the fundamental mechanisms of turbulent transport for bulk ions, impurities, and fast particles, and to explore their relationship with discharge parameters, such as magnetic configurations and turbulence properties.


\section*{Acknowledgements}
This work has been carried out within the framework of the EUROfusion Consortium, funded by the European Union via the Euratom Research and Training Programme (Grant Agreement No 101052200 — EUROfusion). Views and opinions expressed are however those of the author(s) only and do not necessarily reflect those of the European Union or the European Commission. Neither the European Union nor the European Commission can be held responsible for them.

This work was also supported by a grant of the Ministry of Research, Innovation and Digitization, CNCS - UEFISCDI, project number PN-IV-P2-2.1-TE-2023-1102, within PNCDI IV.

\appendix
\section{Collision terms}
\label{A1}

In this appendix, we expand on the formal definitions of the terms that appear in the collisional stochastic components of motion for gyrocenters (\ref{Section_2.3}). More details can be found in literature \cite{HIRVIJOKI20141310,Hirvijoki_monte_carlo_coll}. We restate here some definitions while adding other new ones: $\mu=m v_\perp^2/2B$, $\mathbf{v}=\mathbf{v}_\perp + \mathbf{b}v_\parallel$, $E_k=mv^2/2$, $\lambda = v_\parallel/v$, $\mathbf{b}=\mathbf{B}/B, |\mathbf{B}| = B$, $x_s = v/\sqrt{2T_s/m_s}$, $c_s = e^4Z_s^2\ln\Lambda/(8\pi\varepsilon_0^2 m^2)$

\begin{align}
	D_c &= \frac{1}{2\Omega^2}\left(D_\parallel (1-\lambda^2)+ D_\perp (1+\lambda^2)\right)\\
    D_\parallel &= \sum_s 4n_s c_s\sqrt{m_s/2T_s}\frac{\psi(x)}{x}\\
    D_\perp &= 	 \sum_s 2n_s c_s\sqrt{m_s/2T_s}\left(\phi(x)-\psi(x)\right)\\
    \nu &= \sum_s \frac{4n_sc_sm_s}{T_s}\left(1+\frac{m}{m_s}\right)\psi(x)
\end{align}

\begin{align}
	\phi(x) &= \frac{2}{\sqrt{\pi}}\int_0^x e^{-y^2}dy\\
	\psi(x) &= \frac{\phi(x)-x\phi'(x)}{2x^2}	
	\end{align}

All these are valid for the case when our species of interest suffers collisions with other species $s$ that are at Maxwellian equilibrium and move, also, with the plasma toroidal rotation $\mathbf{u}$. 

The differential Wiener process is, numerically, equivalent with a Gaussian white noise $\chi(t), \langle\chi(t)\rangle = 0, \langle\chi(t)\chi(t^\prime)\rangle = \delta_{t,t^\prime}$, divided by the time step: $d\mathcal{W}^i(t) = \chi(t)/\sqrt{dt}$. 

The matrix $\hat{\Sigma}=\begin{pmatrix}
	\Sigma^{\mu,\mu} & \Sigma^{\mu,v_\parallel} \\
	\Sigma^{v_\parallel,\mu} & \Sigma^{v_\parallel,v_\parallel}
\end{pmatrix}$ obeys the equation:

\begin{align}\hat{\Sigma}\cdot\hat{\Sigma}= \begin{pmatrix}
\frac{E_{kin}}{vB} & 0\\
0 & 1
\end{pmatrix}\cdot\begin{pmatrix}
4(1-\lambda^2)\left[D_\parallel(1-\lambda^2) + D_\perp\lambda^2\right] & 2\lambda(1-\lambda^2)(D_\parallel-D_\perp)\\
2\lambda(1-\lambda^2)(D_\parallel-D_\perp) & D_\parallel \lambda^2 + D_\perp (1-\lambda^2)
\end{pmatrix} \cdot\begin{pmatrix}
\frac{E_{kin}}{vB} & 0\\
0 & 1
\end{pmatrix}
\end{align}

\bibliographystyle{unsrt}
\bibliography{biblio}

\begin{thebibliography}{10}

\bibitem{wesson2011tokamaks}
John Wesson and David~J Campbell.
\newblock {\em Tokamaks}, volume 149.
\newblock Oxford university press, 2011.

\bibitem{Mazzi2022}
S.~Mazzi, et~al, and J.~E.~T. Contributors.
\newblock Enhanced performance in fusion plasmas through turbulence suppression
  by megaelectronvolt ions.
\newblock {\em Nature Physics}, 18(7):776--782, Jul 2022.

\bibitem{Joffrin_2024}
E.~Joffrin, M.~Wischmeier, M.~Baruzzo, A.~Hakola, A.~Kappatou, D.~Keeling,
  B.~Labit, E.~Tsitrone, N.~Vianello, the ASDEX Upgrade~Team, JET Contributors,
  the MAST-U~Team, the TCV~Team, the WEST~Team, and the EUROfusion Tokamak
  Exploitation~Team.
\newblock Overview of the eurofusion tokamak exploitation programme in support
  of iter and demo.
\newblock {\em Nuclear Fusion}, 64(11):112019, aug 2024.

\bibitem{Mailloux_2022}
J.~Mailloux and et~al.
\newblock Overview of jet results for optimising iter operation.
\newblock {\em Nuclear Fusion}, 62(4):042026, jun 2022.

\bibitem{Hirshman_1981}
S.P. Hirshman and D.J. Sigmar.
\newblock Neoclassical transport of impurities in tokamak plasmas.
\newblock {\em Nuclear Fusion}, 21(9):1079--1201, sep 1981.

\bibitem{RevModPhys.71.735}
W.~Horton.
\newblock Drift waves and transport.
\newblock {\em Rev. Mod. Phys.}, 71:735--778, Apr 1999.

\bibitem{Brizard_hahm_gyrokinetic}
A.~J. Brizard and T.~S. Hahm.
\newblock Foundations of nonlinear gyrokinetic theory.
\newblock {\em Rev. Mod. Phys.}, 79:421--468, Apr 2007.

\bibitem{Belli_2008}
E~A Belli and J~Candy.
\newblock Kinetic calculation of neoclassical transport including
  self-consistent electron and impurity dynamics.
\newblock {\em Plasma Physics and Controlled Fusion}, 50(9):095010, jul 2008.

\bibitem{10.1063/1.872465}
W.~A. Houlberg, K.~C. Shaing, S.~P. Hirshman, and M.~C. Zarnstorff.
\newblock Bootstrap current and neoclassical transport in tokamaks of arbitrary
  collisionality and aspect ratio.
\newblock {\em Physics of Plasmas}, 4(9):3230--3242, 09 1997.

\bibitem{HIRVIJOKI20141310}
E.~Hirvijoki, O.~Asunta, T.~Koskela, T.~Kurki-Suonio, J.~Miettunen, S.~Sipilä,
  A.~Snicker, and S.~Äkäslompolo.
\newblock Ascot: Solving the kinetic equation of minority particle species in
  tokamak plasmas.
\newblock {\em Computer Physics Communications}, 185(4):1310--1321, 2014.

\bibitem{LANTI2020107072}
E.~Lanti, N.~Ohana, N.~Tronko, T.~Hayward-Schneider, A.~Bottino, B.F. McMillan,
  A.~Mishchenko, A.~Scheinberg, A.~Biancalani, P.~Angelino, S.~Brunner,
  J.~Dominski, P.~Donnel, C.~Gheller, R.~Hatzky, A.~Jocksch, S.~Jolliet, Z.X.
  Lu, J.P. {Martin Collar}, I.~Novikau, E.~Sonnendrücker, T.~Vernay, and
  L.~Villard.
\newblock Orb5: A global electromagnetic gyrokinetic code using the pic
  approach in toroidal geometry.
\newblock {\em Computer Physics Communications}, 251:107072, 2020.

\bibitem{JENKO2000196}
Frank Jenko.
\newblock Massively parallel vlasov simulation of electromagnetic drift-wave
  turbulence.
\newblock {\em Computer Physics Communications}, 125(1):196--209, 2000.

\bibitem{PEETERS20092650}
A.G. Peeters, Y.~Camenen, F.J. Casson, W.A. Hornsby, A.P. Snodin, D.~Strintzi,
  and G.~Szepesi.
\newblock The nonlinear gyro-kinetic flux tube code gkw.
\newblock {\em Computer Physics Communications}, 180(12):2650--2672, 2009.
\newblock 40 YEARS OF CPC: A celebratory issue focused on quality software for
  high performance, grid and novel computing architectures.

\bibitem{CANDY2003545}
J.~Candy and R.E. Waltz.
\newblock An eulerian gyrokinetic-maxwell solver.
\newblock {\em Journal of Computational Physics}, 186(2):545--581, 2003.

\bibitem{GRANDGIRARD2006395}
V.~Grandgirard, M.~Brunetti, P.~Bertrand, N.~Besse, X.~Garbet, P.~Ghendrih,
  G.~Manfredi, Y.~Sarazin, O.~Sauter, E.~Sonnendrücker, J.~Vaclavik, and
  L.~Villard.
\newblock A drift-kinetic semi-lagrangian 4d code for ion turbulence
  simulation.
\newblock {\em Journal of Computational Physics}, 217(2):395--423, 2006.

\bibitem{Idomura_2005}
Y.~Idomura, S.~Tokuda, and Y.~Kishimoto.
\newblock Global profile effects and structure formations in toroidal electron
  temperature gradient driven turbulence.
\newblock {\em Nuclear Fusion}, 45(12):1571, nov 2005.

\bibitem{Bourdelle_2016}
C~Bourdelle, J~Citrin, B~Baiocchi, A~Casati, P~Cottier, X~Garbet, F~Imbeaux,
  and JET Contributors.
\newblock Core turbulent transport in tokamak plasmas: bridging theory and
  experiment with qualikiz.
\newblock {\em Plasma Physics and Controlled Fusion}, 58(1):014036, nov 2015.

\bibitem{Stephens_Garbet_2021}
C.D. Stephens, X.~Garbet, J.~Citrin, C.~Bourdelle, K.L. van~de Plassche, and
  F.~Jenko.
\newblock Quasilinear gyrokinetic theory: a derivation of qualikiz.
\newblock {\em Journal of Plasma Physics}, 87(4):905870409, 2021.

\bibitem{10.1063/5.0174643}
E.~Fransson, A.~Gillgren, A.~Ho, J.~Borsander, O.~Lindberg, W.~Rieck,
  M.~Åqvist, and P.~Strand.
\newblock A fast neural network surrogate model for the eigenvalues of
  qualikiz.
\newblock {\em Physics of Plasmas}, 30(12):123904, 12 2023.

\bibitem{10.1063/1.2800869}
C.~Bourdelle, X.~Garbet, F.~Imbeaux, A.~Casati, N.~Dubuit, R.~Guirlet, and
  T.~Parisot.
\newblock A new gyrokinetic quasilinear transport model applied to particle
  transport in tokamak plasmas.
\newblock {\em Physics of Plasmas}, 14(11):112501, 11 2007.

\bibitem{Ward_2021}
S.H. Ward, R.~Akers, A.S. Jacobsen, P.~Ollus, S.D. Pinches, E.~Tholerus, R.G.L.
  Vann, and M.A. Van~Zeeland.
\newblock Verification and validation of the high-performance lorentz-orbit
  code for use in stellarators and tokamaks (locust).
\newblock {\em Nuclear Fusion}, 61(8):086029, jul 2021.

\bibitem{osti_5168290}
R~B White and M~S Chance.
\newblock Hamiltonian guiding center drift orbit calculation for toroidal
  plasmas of arbitrary cross section.
\newblock Technical report, Princeton Plasma Physics Lab. (PPPL), Princeton, NJ
  (United States), 02 1984.

\bibitem{Wang_2021}
Feng Wang, Rui Zhao, Zheng-Xiong Wang, Yue Zhang, Zhan-Hong Lin, Shi-Jie Liu,
  and CFETR Team.
\newblock Ptc: Full and drift particle orbit tracing code for $\alpha$
  particles in tokamak plasmas.
\newblock {\em Chinese Physics Letters}, 38(5):055201, jun 2021.

\bibitem{Palade2021}
D.~I. Palade and M.~Vlad.
\newblock Fast generation of gaussian random fields for direct numerical
  simulations of stochastic transport.
\newblock {\em Statistics and Computing}, 31(5):60, Aug 2021.

\bibitem{Palade_2021_W}
Dragos~Iustin Palade, Madalina Vlad, and Florin Spineanu.
\newblock Turbulent transport of the w ions in tokamak plasmas: properties
  derived from a test particle approach.
\newblock {\em Nuclear Fusion}, 61(11):116031, oct 2021.

\bibitem{Palade_2023}
D.I. Palade.
\newblock Peaking and hollowness of low-z impurity profiles: an interplay
  between itg and tem induced turbulent transport.
\newblock {\em Nuclear Fusion}, 63(4):046007, mar 2023.

\bibitem{Vlad_2021}
Madalina Vlad, Dragos~Iustin Palade, and Florin Spineanu.
\newblock Effects of the parallel acceleration on heavy impurity transport in
  turbulent tokamak plasmas.
\newblock {\em Plasma Physics and Controlled Fusion}, 63(3):035007, jan 2021.

\bibitem{palade_pom_2022}
D.I. Palade and L.~Pomârjanschi.
\newblock Effects of intermittency via non-gaussianity on turbulent transport
  in magnetized plasmas.
\newblock {\em Journal of Plasma Physics}, 88(2):905880202, 2022.

\bibitem{Palade_2024_scaling}
D~I Palade, L~M Pomârjanschi, and M~Ghiţă.
\newblock Scaling laws of two-dimensional incompressible turbulent transport.
\newblock {\em Physica Scripta}, 99(1):015201, dec 2023.

\bibitem{balescu2005aspects}
Radu Balescu.
\newblock {\em Aspects of anomalous transport in plasmas}.
\newblock CRC Press, 2005.

\bibitem{10.1063/1.1647136}
Z.~Lin and T.~S. Hahm.
\newblock Turbulence spreading and transport scaling in global gyrokinetic
  particle simulations.
\newblock {\em Physics of Plasmas}, 11(3):1099--1108, 03 2004.

\bibitem{10.1063/1.4844035}
M.~Vlad and F.~Spineanu.
\newblock Test particle study of ion transport in drift type turbulence.
\newblock {\em Physics of Plasmas}, 20(12):122304, 12 2013.

\bibitem{PhysRevLett.91.035001}
X.~Garbet, L.~Garzotti, P.~Mantica, H.~Nordman, M.~Valovic, H.~Weisen, and
  C.~Angioni.
\newblock Turbulent particle transport in magnetized plasmas.
\newblock {\em Phys. Rev. Lett.}, 91:035001, Jul 2003.

\bibitem{PhysRevLett.88.195004}
Z.~Lin, S.~Ethier, T.~S. Hahm, and W.~M. Tang.
\newblock Size scaling of turbulent transport in magnetically confined plasmas.
\newblock {\em Phys. Rev. Lett.}, 88:195004, Apr 2002.

\bibitem{PhysRevLett.101.095001}
Wenlu Zhang, Zhihong Lin, and Liu Chen.
\newblock Transport of energetic particles by microturbulence in magnetized
  plasmas.
\newblock {\em Phys. Rev. Lett.}, 101:095001, Aug 2008.

\bibitem{10.1063/1.3379471}
Wenlu Zhang, Viktor Decyk, Ihor Holod, Yong Xiao, Zhihong Lin, and Liu Chen.
\newblock Scalings of energetic particle transport by ion temperature gradient
  microturbulence.
\newblock {\em Physics of Plasmas}, 17(5):055902, 05 2010.

\bibitem{Vlad_2004}
M~Vlad, F~Spineanu, J~H Misguich, J-D Reuss, R~Balescu, K~Itoh, and S-I Itoh.
\newblock Lagrangian versus eulerian correlations and transport scaling.
\newblock {\em Plasma Physics and Controlled Fusion}, 46(7):1051--1063, may
  2004.

\bibitem{10.1063/1.3013453}
T.~Hauff and F.~Jenko.
\newblock Mechanisms and scalings of energetic ion transport via tokamak
  microturbulence.
\newblock {\em Physics of Plasmas}, 15(11):112307, 11 2008.

\bibitem{hauff0}
T.~Hauff and F.~Jenko.
\newblock {Turbulent E×B advection of charged test particles with large
  gyroradii}.
\newblock {\em Physics of Plasmas}, 13(10):102309, 10 2006.

\bibitem{hauff1}
T.~Hauff and F.~Jenko.
\newblock {E×B advection of trace ions in tokamak microturbulence}.
\newblock {\em Physics of Plasmas}, 14(9):092301, 09 2007.

\bibitem{Camenen_rotation}
Y.~Camenen, A.~G. Peeters, C.~Angioni, F.~J. Casson, W.~A. Hornsby, A.~P.
  Snodin, and D.~Strintzi.
\newblock {Impact of the background toroidal rotation on particle and heat
  turbulent transport in tokamak plasmas}.
\newblock {\em Physics of Plasmas}, 16(1):012503, 01 2009.

\bibitem{sugama2017modern}
H~Sugama.
\newblock Modern gyrokinetic formulation of collisional and turbulent transport
  in toroidally rotating plasmas.
\newblock {\em Reviews of Modern Plasma Physics}, 1(1):9, 2017.

\bibitem{brizard_rotating}
Alain~J. Brizard.
\newblock {Nonlinear gyrokinetic Vlasov equation for toroidally rotating
  axisymmetric tokamaks}.
\newblock {\em Physics of Plasmas}, 2(2):459--471, 02 1995.

\bibitem{wang_hahm_polarization}
Lu~Wang and T.~S. Hahm.
\newblock {Nonlinear gyrokinetic theory with polarization drift}.
\newblock {\em Physics of Plasmas}, 17(8):082304, 08 2010.

\bibitem{Littlejohn}
Robert~G. Littlejohn.
\newblock {Hamiltonian perturbation theory in noncanonical coordinates}.
\newblock {\em Journal of Mathematical Physics}, 23(5):742--747, 05 1982.

\bibitem{SAUTER2013293}
O.~Sauter and S.Yu. Medvedev.
\newblock Tokamak coordinate conventions: Cocos.
\newblock {\em Computer Physics Communications}, 184(2):293--302, 2013.

\bibitem{Xanthopoulos}
P.~Xanthopoulos and F.~Jenko.
\newblock {Clebsch-type coordinates for nonlinear gyrokinetics in generic
  toroidal configurations}.
\newblock {\em Physics of Plasmas}, 13(9):092301, 09 2006.

\bibitem{GORLER20117053}
T.~Görler, X.~Lapillonne, S.~Brunner, T.~Dannert, F.~Jenko, F.~Merz, and
  D.~Told.
\newblock The global version of the gyrokinetic turbulence code gene.
\newblock {\em Journal of Computational Physics}, 230(18):7053--7071, 2011.

\bibitem{Beer_fieldaligned}
M.~A. Beer, S.~C. Cowley, and G.~W. Hammett.
\newblock {Field‐aligned coordinates for nonlinear simulations of tokamak
  turbulence}.
\newblock {\em Physics of Plasmas}, 2(7):2687--2700, 07 1995.

\bibitem{Hinton_neoclassical}
F.~L. Hinton and S.~K. Wong.
\newblock {Neoclassical ion transport in rotating axisymmetric plasmas}.
\newblock {\em The Physics of Fluids}, 28(10):3082--3098, 10 1985.

\bibitem{Peeters_rotation}
A.~G. Peeters, D.~Strintzi, Y.~Camenen, C.~Angioni, F.~J. Casson, W.~A.
  Hornsby, and A.~P. Snodin.
\newblock {Influence of the centrifugal force and parallel dynamics on the
  toroidal momentum transport due to small scale turbulence in a tokamak}.
\newblock {\em Physics of Plasmas}, 16(4):042310, 04 2009.

\bibitem{10.1063/1.1694506}
Thomas~H. Stix.
\newblock {Decay of poloidal rotation in a tokamak plasma}.
\newblock {\em The Physics of Fluids}, 16(8):1260--1267, 08 1973.

\bibitem{Evans2006}
et~al Evans, Todd~E.
\newblock Edge stability and transport control with resonant magnetic
  perturbations in collisionless tokamak plasmas.
\newblock {\em Nature Physics}, 2(6):419--423, Jun 2006.

\bibitem{Merz_2010}
F.~Merz and F.~Jenko.
\newblock Nonlinear interplay of tem and itg turbulence and its effect on
  transport.
\newblock {\em Nuclear Fusion}, 50(5):054005, apr 2010.

\bibitem{palade_2025_inhomogeneous}
D.~I. Palade.
\newblock {Nonlinear transport coefficients in inhomogeneous magnetized plasmas
  (via particle trajectories)}.
\newblock {\em work in progress}.

\bibitem{shafranov1958magnetohydrodynamical}
VD~Shafranov.
\newblock On magnetohydrodynamical equilibrium configurations.
\newblock {\em Soviet Physics JETP}, 6(3):1013, 1958.

\bibitem{Atanasiu}
C.~V. Atanasiu, S.~Günter, K.~Lackner, and I.~G. Miron.
\newblock {Analytical solutions to the Grad–Shafranov equation}.
\newblock {\em Physics of Plasmas}, 11(7):3510--3518, 07 2004.

\bibitem{solov1968theory}
LS~Solov’ev.
\newblock The theory of hydromagnetic stability of toroidal plasma
  configurations.
\newblock {\em Sov. Phys. JETP}, 26(2):400--407, 1968.

\bibitem{LUTJENS1996219}
H.~Lütjens, A.~Bondeson, and O.~Sauter.
\newblock The chease code for toroidal mhd equilibria.
\newblock {\em Computer Physics Communications}, 97(3):219--260, 1996.

\bibitem{Lao_1985}
L.L. Lao, H.~St. John, R.D. Stambaugh, A.G. Kellman, and W.~Pfeiffer.
\newblock Reconstruction of current profile parameters and plasma shapes in
  tokamaks.
\newblock {\em Nuclear Fusion}, 25(11):1611, nov 1985.

\bibitem{kripner2019towards}
L~Kripner, M~Tome{\v{s}}, M~Peterka, J~Urban, E~Mac{\'u}{\v{s}}ov{\'a},
  F~Jaulmes, D~Fridrich, O~Grover, O~Ficker, J~Krbec, et~al.
\newblock Towards the integrated analysis of tokamak plasma equilibria: Pleque.
\newblock In {\em 46th EPS Conf. on Plasma Physics}, 2019.

\bibitem{FAUGERAS2020112020}
Blaise Faugeras.
\newblock An overview of the numerical methods for tokamak plasma equilibrium
  computation implemented in the nice code.
\newblock {\em Fusion Engineering and Design}, 160:112020, 2020.

\bibitem{BLUM2012960}
J.~Blum, C.~Boulbe, and B.~Faugeras.
\newblock Reconstruction of the equilibrium of the plasma in a tokamak and
  identification of the current density profile in real time.
\newblock {\em Journal of Computational Physics}, 231(3):960--980, 2012.
\newblock Special Issue: Computational Plasma Physics.

\bibitem{Brizard_collision}
Alain~J. Brizard.
\newblock {A guiding-center Fokker–Planck collision operator for nonuniform
  magnetic fields}.
\newblock {\em Physics of Plasmas}, 11(9):4429--4438, 09 2004.

\bibitem{Hirvijoki_monte_carlo_coll}
E.~Hirvijoki, A.~Brizard, A.~Snicker, and T.~Kurki-Suonio.
\newblock {Monte Carlo implementation of a guiding-center Fokker-Planck kinetic
  equation}.
\newblock {\em Physics of Plasmas}, 20(9):092505, 09 2013.

\bibitem{boozer1980monte}
Allen~H Boozer and Gioietta Kuo-Petravic.
\newblock Monte carlo evaluation of transport coefficients.
\newblock Technical report, Princeton Plasma Physics Lab.(PPPL), Princeton, NJ
  (United States), 1980.

\bibitem{PhysRevLett.85.5579}
W.~Dorland, F.~Jenko, M.~Kotschenreuther, and B.~N. Rogers.
\newblock Electron temperature gradient turbulence.
\newblock {\em Phys. Rev. Lett.}, 85:5579--5582, Dec 2000.

\bibitem{Evans_2015}
T~E Evans.
\newblock Resonant magnetic perturbations of edge-plasmas in toroidal
  confinement devices.
\newblock {\em Plasma Physics and Controlled Fusion}, 57(12):123001, nov 2015.

\bibitem{PhysRevLett.40.396}
J.~W. Connor, R.~J. Hastie, and J.~B. Taylor.
\newblock Shear, periodicity, and plasma ballooning modes.
\newblock {\em Phys. Rev. Lett.}, 40:396--399, Feb 1978.

\bibitem{10.1063/1.4954905}
G.~M. Staebler, J.~Candy, N.~T. Howard, and C.~Holland.
\newblock The role of zonal flows in the saturation of multi-scale gyrokinetic
  turbulence.
\newblock {\em Physics of Plasmas}, 23(6):062518, 06 2016.

\bibitem{Dudding_2022}
H.G. Dudding, F.J. Casson, D.~Dickinson, B.S. Patel, C.M. Roach, E.A. Belli,
  and G.M. Staebler.
\newblock A new quasilinear saturation rule for tokamak turbulence with
  application to the isotope scaling of transport.
\newblock {\em Nuclear Fusion}, 62(9):096005, jul 2022.

\bibitem{10.1063/1.859070}
S.~J. Zweben and S.~S. Medley.
\newblock Visible imaging of edge fluctuations in the tftr tokamak.
\newblock {\em Physics of Fluids B: Plasma Physics}, 1(10):2058--2065, 10 1989.

\bibitem{1944519}
Kiyosi ITO.
\newblock 109. stochastic integral.
\newblock {\em Proceedings of the Imperial Academy}, 20(8):519--524, 1944.

\bibitem{palade2020fast}
DI~Palade and M~Vlad.
\newblock Fast generation of gaussian random fields for direct numerical
  simulations of stochastic transport.
\newblock {\em arXiv preprint arXiv:2006.11106 (submitted)}, 2020.

\bibitem{A.M.Dimits_2000}
A.M. Dimits, B.I. Cohen, N.~Mattor, W.M. Nevins, D.E. Shumaker, S.E. Parker,
  and C.~Kim.
\newblock Simulation of ion temperature gradient turbulence in tokamaks.
\newblock {\em Nuclear Fusion}, 40(3Y):661, mar 2000.

\bibitem{Falchetto_2008}
G~L Falchetto, B~D Scott, P~Angelino, A~Bottino, T~Dannert, V~Grandgirard,
  S~Janhunen, F~Jenko, S~Jolliet, A~Kendl, B~F McMillan, V~Naulin, A~H Nielsen,
  M~Ottaviani, A~G Peeters, M~J Pueschel, D~Reiser, T~T Ribeiro, and
  M~Romanelli.
\newblock The european turbulence code benchmarking effort: turbulence driven
  by thermal gradients in magnetically confined plasmas.
\newblock {\em Plasma Physics and Controlled Fusion}, 50(12):124015, nov 2008.

\bibitem{Dong_2022}
G~Dong and Z~Lin.
\newblock Role of wave-particle resonance in turbulent transport in toroidal
  plasmas.
\newblock {\em Plasma Physics and Controlled Fusion}, 64(3):035005, jan 2022.

\bibitem{Pueschel_2012}
M.J. Pueschel, F.~Jenko, M.~Schneller, T.~Hauff, S.~Günter, and G.~Tardini.
\newblock Anomalous diffusion of energetic particles: connecting experiment and
  simulations.
\newblock {\em Nuclear Fusion}, 52(10):103018, sep 2012.

\end{thebibliography}

\end{document}